# Novel λ³-Iodane Based Functionalization of Synthetic Carbon Allotropes (SCAs) – Common Concepts and Quantification of the Degree of Addition

Ferdinand Hof, Ricarda A. Schäfer, Cornelius Weiss, Frank Hauke and Andreas Hirsch*[a]

F. Hof, R. A. Schäfer, C. Weiss, Dr. F. Hauke and Prof. Dr. A. Hirsch
Department of Chemistry and Pharmacy and Institute of Advanced Materials and Processes (ZMP)
Friedrich-Alexander-Universität Erlangen-Nürnberg (FAU)
Henkestrasse 42, 91054 Erlangen, Germany
Fax: (+) 49 (0)9131 85 26864
E-mail: andreas.hirsch@fau.de

**Abstract:** The covalent functionalization of carbon allotropes represents a main topic in the growing field of nano materials. However, the development of functional architectures is impeded by the intrinsic polydispersibility of the respective starting material, the unequivocal characterization of the introduced functional moieties and the exact determination of the degree of functionalization. Based on a novel carbon allotrope functionalization reaction, utilizing λ³-iodanes as radical precursor systems, we were able to demonstrate the feasibility to separate and to quantify thermally detached functional groups, formerly covalently linked to carbon nanotubes and graphene via TG-GC/MS.

## Introduction

Inspired by the rich chemistry of fullerenes, the functionalization of other synthetic carbon allotropes (SCAs), in particular carbon nanotubes and more recently graphene, has steadily been progressed.[1] In parallel to the development of efficient reaction protocols, such as the reductive alkylation and arylation of SCAs,[2] tools for product characterization have continuously been adjusted and improved.[3] Typical techniques are: thermogravimetric analysis,[4] Raman spectroscopy,[3g] STM/AFM,[5] and XPS.[6] However, the powerful analytical tool-kit, routinely used for the structural characterization of organic molecules, such as NMR spectroscopy, mass spectrometry and X-ray crystallography can hardly be applied for the analysis of covalently functionalized carbon nanotubes and graphene, due to their intrinsic polydisperse nature. Here, only mixtures of adducts with a broad distribution of sizes, shapes and degrees of addition are accessible for analysis. Therefore, only averaging information can be obtained, which requires the development/application of techniques for data acquisition founded on a broad statistical basis. Along these lines, we have recently introduced scanning Raman spectroscopy (SRS) and –microscopy (SRM) as versatile techniques for a reliable and reproducible characterization of covalently functionalized SCA bulk samples.[3f, 3g] Despite these improvements, there are still fundamental questions that have not been addressed or solved so far: a) the exact quantification of the degree of addition of a specific functional entity and b) a direct comparison of the reactivity of different SCAs under equivalent reaction conditions. Herein, we provide a profound answer for these two fundamental challenges. Based on a novel and versatile reductive arylation of graphite/graphene and carbon nanotubes – utilizing $\lambda^3$-iodane precursors – the degree of functionalization has been quantified by a thermo gravimetric product analysis, coupled with gas-chromatographic separation and mass spectrometric characterization (TG-GC/MS). With this setup, we are now able to separate the thermally detached functional entities and to selectively identify and quantify each individual component attached to the carbon allotrope framework.

**Results and Discussion**

For the $\lambda^3$-iodane based reductive arylation, the respective SCA was initially reduced with a stoichiometric amount of potassium yielding the corresponding nanotubide and graphenide salts,[7] which represent highly activated intermediates. We have shown recently, that the stoichiometry of potassium is a critical factor to fine tune the degree of addition in the final SWCNT derivatives.[8] Therefore, the investigation of the reaction of nanotubides and graphenides with different $\lambda^3$-iodanes was carried out by the variation of the potassium:carbon ratio (**T$_{(1:4)}$**, **T$_{(1:8)}$**, **T$_{(1:16)}$**, **T$_{(1:24)}$** and **G$_{(1:4)}$**, **G$_{(1:8)}$**, **G$_{(1:16)}$**, **G$_{(1:24)}$** exhibiting a respective carbon to potassium ration of 1:4, 1:8, 1:16 and 1:24) (Scheme 1).

Hypervalent iodine compounds easily form radicals and can be used as functional group transfer reagents.[9] In SCA chemistry they have so far only been applied for the electrochemical functionalization of carbon electrodes[10] as well as other types of surfaces.[11] In line with the radical mechanisms stated for reductive arylation/alkylation reactions and diazonium based SCA functionalizations,[12] an equivalent pathway can be anticipated for the reaction of $\lambda^3$-iodanes. In order to screen their reactivity as SCA trapping electrophiles, three different para functionalized $\lambda^3$-iodanes *bis*-(4-(t-butyl)phenyl) iodonium trifluoromethane-sulfonate **A**, (4-bromophenyl)(4-(t-butyl)phenyl iodonium trifluoro-methanesulfonate **B**, and *bis*-(4-bromophenyl) iodonium trifluoro-methanesulfonate **C** have been synthesized by adjusting literature known procedures (for details see ESI).[13] For the synthesis of the arylated SCA derivatives, the corresponding potassium nanotubide/graphenide salt was dispersed in THF$_{abs}$ (glove box: Ar) and subsequently 0.5 equivalents per carbon of the respective $\lambda^3$-iodane **A-C** was added. The isolation of the reaction products **T(A-C)** and **G(A-C)** was accomplished by aqueous workup and filtration. In order to gain a statistically significant bulk information about the success of this covalent functionalization pathway scanning Raman spectroscopy (SRS) was carried out. Here, the $I_{(D/G)}$ (intensity ratios of the D- and G-bands) values of the individual Raman spectra (at least 1.250 single point spectra measured) are plotted with respect to their frequency and the respective distribution function was determined (Figure 1).

In principle, the D-band intensity is a measure for the amount of lattice sp³-centers introduced by covalent addend binding. The successful reaction of the reduced SCAs with the λ³-iodanes **A-C** can exemplarily be demonstrated (K:C = 1:4) on the basis of the Raman data of the functionalized SWCNTs ($\lambda_{exc}$ = 633 nm – Fig 1, top) and the graphite/graphene[14] derivatives ($\lambda_{exc}$ = 532 nm – Fig. 1, bottom) – Table 1. For further Raman data see ESI: Fig. S1, S2.

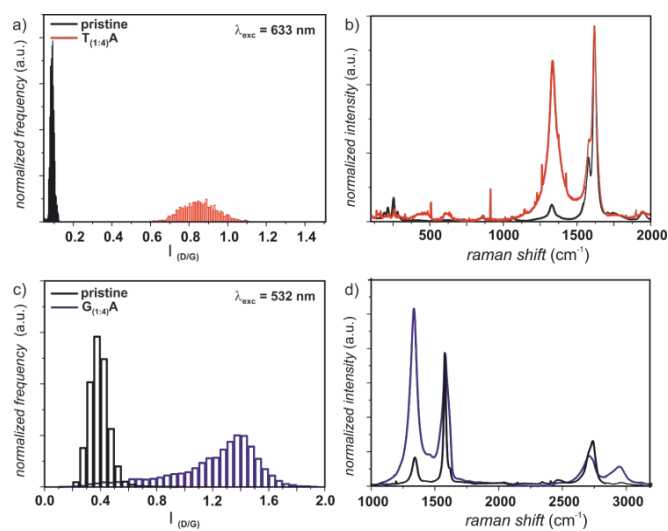

**Figure 1.** Statistical Raman spectroscopic analysis of the functionalized carbon allotrope **T$_{(1:4)}$A** ($\lambda_{exc}$ = 633 nm) and **G$_{(1:4)}$A** ($\lambda_{exc}$ = 532 nm). Top: Histogram and mean spectrum **T$_{(1:4)}$A**. Bottom: Histogram and mean spectrum of **G$_{(1:4)}$A**.

The potential of this novel synthetic route becomes apparent, when the respective $I_{(D/G)}$ values of the carbon nanotube samples **T$_{(1:4)}$A,B,C** are compared with reference samples prepared by literature known synthetic pathways obtained by reductive hexylation **T$_{(1:4)}$Hex-I** or reductive arylation **T$_{(1:4)}$Ph-I** of carbon nanotubes.[15] – Fig. 2.

**Table 1.** Statistical Raman spectroscopic data of the reaction products **T(1:n)A,B,C** and **G(1:n)A,B,C** - $I_{(D/G)}$ values with standard deviation Δ.

| SWCNT | $I_{(D/G)}$ 633 nm | Δ | graphite/ graphene | $I_{(D/G)}$ 532 nm | Δ |
|---|---|---|---|---|---|
| **T$_{(1:4)}$A** | 0.95 | 0.15 | **G$_{(1:4)}$A** | 1.00 | 0.30 |
| **T$_{(1:4)}$B** | 0.90 | 0.11 | **G$_{(1:4)}$B** | 1.05 | 0.30 |
| **T$_{(1:4)}$C** | 0.93 | 0.17 | **G$_{(1:4)}$C** | 0.95 | 0.30 |
| **T$_{(1:8)}$A** | 0.68 | 0.17 | **G$_{(1:8)}$A** | 1.25 | 0.11 |
| **T$_{(1:16)}$A** | 0.42 | 0.24 | **G$_{(1:16)}$A** | 1.05 | 0.30 |
| **T$_{(1:24)}$A** | 0.26 | 0.09 | **G$_{(1:24)}$A** | 0.81 | 0.47 |

For arylated SWCNT derivatives accessible by our novel λ³-iodane based functionalization sequence, the degrees of functional group addition, represented by the respective RDI indices, drastically exceed the values obtained for systems synthesized by classical pathways[15]: *e.g.* RDI(**T(1:4)A**) = 1.83 *vs.* RDI(**T(1:4)Ph-I**) = 0.83. In accordance with our latest results[8] we also observe a pronounced dependence of the Raman derived $I_{(D/G)}$ ratio on the amount of potassium used for the reduction for the SWCNTs samples and for the graphite/graphene samples (Figure S4-S6; Table 1). The final degree of functionalization increases with the respective amount of potassium used in the reductive SCA activation step.

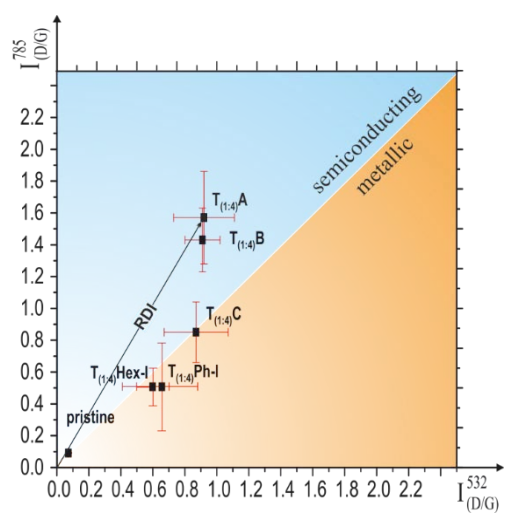

**Figure 2.** 2D Raman index plot based on the statistical Raman analysis of **T(1:4)A**, **T(1:4)B**, **T(1:4)C** and of the alkylated and arylated reference samples **T(1:4)Hex-I** and **T(1:4)Ph-I**. The respective values for the Raman Defect Index (RDI), Raman Homogeneity Index (RHI), and Raman Selectivity Index (RSI) are summarized in Fig. S3.

Furthermore, we were able to show that these covalently bound aryl moieties can reversibly be detached from the carbon allotrope framework either by laser induced de-functionalization (exemplarily shown for **T(4:1)A**; $\lambda_{exc}$ = 785 nm, E = 10 mW – Fig. 3 left, Fig. S7) or thermally. The latter approach opens the possibility to monitor the detachment progress of the functional moieties directly by Raman spectroscopy (Fig. 2 right, Fig. S8).

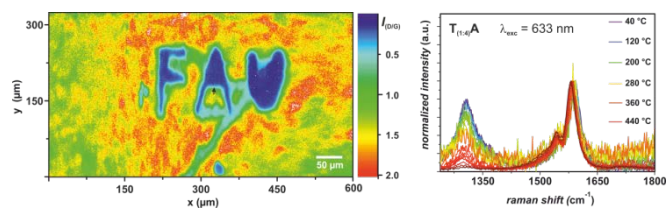

**Figure 3.** Left: Laser induced thermal de-functionalization of the functionalized bulk material **T(1:4)A** demonstrating the reversibility of addend binding. Regions in blue color denote areas with lower degrees of functionalization based on a thermal de-functionalization by "writing" with high laser power. Right: Temperature dependent Raman spectroscopy of *t*-butylphenyl functionalized SWCNTs.

Here, for both functionalized SCA derivatives – $T_{(1:4)}A$ and $G_{(1:4)}A$ – a continuous decrease of D-band intensity is observed starting from around 150 °C up to 500 °C. At temperatures around 500 °C, the characteristic Raman spectrum of the respective starting material is restored exhibiting an $I_{(D/G)}$ value of less than 0.1 in the bulk (Fig. S8). However, neither by SRS nor by temperature depending Raman spectroscopy any conclusion about the chemical nature of the attached moieties can be extracted. Thus, thermogravimetric analysis coupled to mass spectrometry (TG-MS) was carried out (20 - 700 °C, constant flow of $N_2$ (70 mL/min – Fig. S9-S14). In the region between 20 and 130 °C, mainly physisorbed THF and water is released from the samples $T_{(1:4)}A,B,C$. According to the respective mass traces, sample mass loss beyond 500 °C mainly can be traced back to the degradation of the carbon framework. The predominant mass loss of 23.2 % ($T_{(1:4)}A$), 17.9 % ($T_{(1:4)}B$) and 28.0 % ($T_{(1:4)}C$), respectively, is detected between 130 and 500 °C and nicely correlates with the respective flux of the detected ion currents, which are the characteristic fragments originating from the detached *t*-butylbenzene and/or bromobenzene moieties (Fig. S9-S11). The same consistent data-set was obtained for the functionalized graphite/graphene derivatives $G_{(1:4)}A,B,C$. However, in this case the observed overall mass loss is considerably lower: 5.4 % ($G_{(1:4)}A$), 10.5 % ($G_{(1:4)}B$) and 3.2 % ($G_{(1:4)}C$) (see Fig. S12-S14). The TG-MS based data nicely corroborates the Raman results regarding an efficient covalent addend binding in both carbon allotropes. Nevertheless, up to now it was not possible to determine an exact value of the degree of the functionalization of covalently modified SCAs as Raman spectroscopy allows only to draw information about the amount of symmetry breaking defects in the $sp^2$ carbon lattice and TG measurements provides only information about the overall sample mass loss attributed to a thermal detachment of all physisorbed and chemisorbed species. To circumvent these obstacles we employed thermogravimetric based, gas-chromatography coupled to mass spectrometry (TG-GC/MS) - Fig.4. This allowed us to determine the mass-loss region, where the covalently bound addend is detached – based on this data an exact quantification of the amount of bound functional groups is possible.

For this purpose, 1 mL of the carrier gas was collected at a sample temperature of 250 °C (highest mass flux) and the individual components were separated by a high performance GC-column (Elite-5MS capillary column) - temperature gradient: 10 K/min; T = 40 - 240 °C. Subsequently, the respective sample fractions were analyzed by mass spectrometry. The corresponding retention times (RT) are depicted in Fig. 4, right. Here, the bromophenyl moiety of $T_{(1:4)}B$ and $T_{(1:4)}C$ is detected at a retention time (RT) of 4.93 min, whereas the *t*-butylphenyl addend in $T_{(1:4)}A$ and $T_{(1:4)}B$ is detected at RT = 5.84 min – MS fragmentation pattern see Fig. S15. In the mixed functionalized systems $T_{(1:4)}B/G_{(1:4)}B$ both thermally detached functional moieties can be identified, exhibiting a peak area ratio of almost 1:1 (Fig. S16, S17).

This result clearly highlights the potential and versatility of our novel functionalization sequence. Staring from an easily accessible $\lambda^3$-iodane precursor system with two different aryl moieties, carbon allotrope based architectures with complementary functional groups – like for instance donor/acceptor systems – can be build up in an one-pot synthesis.

Next to the identification of the main components by TG-GC/MS, formation of byproducts can also be investigated by this setup. Here, in all three functionalized SWCNT samples (Fig. S16), additional components (below 3 % total peak area) are detected by GC (RT = 1.93, 1.95, 2.04 min), which can be attributed to THF (solvent) and degradation products thereof (furane, dihydrofurane). Remarkably, in the case of the graphite/graphene adducts **G$_{(1:4)}$A,B,C** the peak for THF (RT = 2.04 min) is observed as main component (~80 %) (Fig. S17). Moreover, in direct comparison, the peak area of the detached *t*-butylphenyl moieties for SWCNT derivatives **T$_{(1:4)}$A** and **T$_{(1:4)}$B** is by two orders of magnitude higher than that detected for the corresponding graphite/graphene adducts **G$_{(1:4)}$A** and **G$_{(1:4)}$B**. Apparently, the degree of covalent functionalization is considerably higher for carbon nanotubes. These results can be attributed to the fact, that in the case of graphite, the reduction and dispersing steps do not exclusively yield completely exfoliated and individualized graphene sheets. The detection of THF as major component is likely to be traced back to intercalated solvent molecules, which are trapped between the carbon sheets. This also reduces the accessible carbon surface area for the hypervalent iodine compounds, which decreases the overall degree of functionalization and yields an explanation for the high amount of THF in the TG-GC/MS experiments. This rationale could be corroborated by a reference experiment where a graphite starting material with initial smaller flake size was chosen. Here, only a negligible amount of THF in relation to the *t*-butylphenyl signal is observed by GC (Fig. S18).

In order to confirm that the obtained GC/MS traces originate from covalently bound functional groups and not from intercalated $\lambda^3$-iodanes, we carried out an additional reference experiment where the hypervalent iodine compound **B** was blended with graphite (Fig. 5). The GC elugram of the reference sample differs distinctively from the elugrams of the covalently functionalized SCA derivatives as it contains not only the characteristic fragmentation pattern of the *t*-butyl- and bromophenyl groups, but also several other compounds originating from the trifluoromethyl sulfonate counter ion of the $\lambda^3$-iodane **B** (not found for covalently functionalized samples).

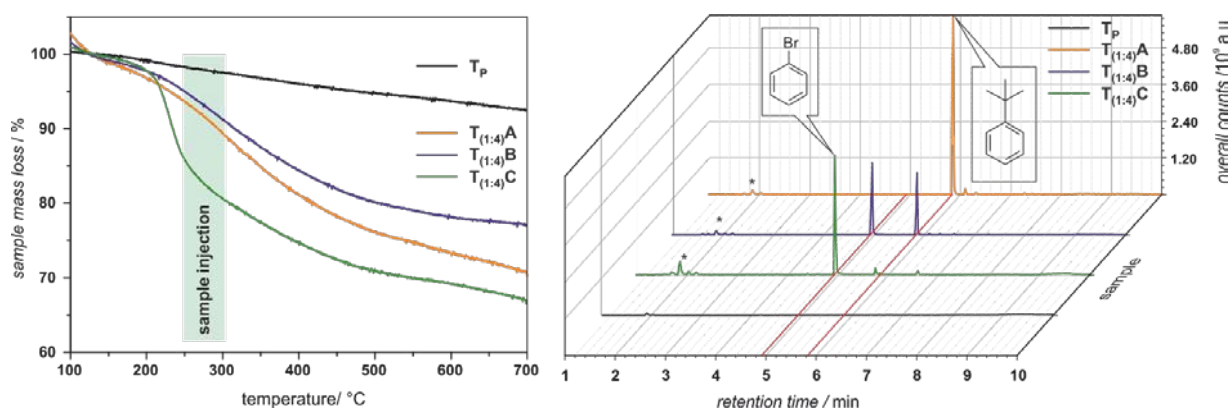

**Figure 4.** TG-GC/MS analysis of the functionalized SWCNT samples $T_{(1:4)}A$, $T_{(1:4)}B$, $T_{(1:4)}C$. a) Left: Plot of mass loss *vs.* temperature. Right: GC/MS analysis of 1 mL gaseous addends detached at 250 °C.

The separation of thermally detached functional moieties and the identification of their chemical identity is a fundamental advantage provided by the TG-GC/MS setup. This fact is also nicely illustrated by the characterization of a *n*-hexyl functionalized SWCNT reference sample (Fig. S19).

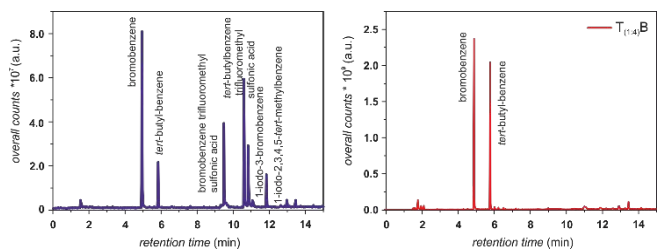

**Figure 5.** Reference experiment – graphite/$\lambda^3$-iodane blend. Left: Elugram of the pristine $\lambda^3$-iodane **B** (TTG = 230 °C) thermal fragmentation products. The mass spectrometric identified compounds are attributed to the respective retention times: 4.93, 5.84, 9.51, 10.62, 10.85, 11.86 min. Right: For comparison - elugram of the covalently functionalized SWCNT derivative $T_{(1:4)}B$.

Here, besides *n*-hexane as main component also the characteristic by-products[8] like hexanol, hexanal and hexyliodide are individually separated by GC and identified by MS. These results nicely confirm our published data concerning the multifunctional product distribution of reductively functionalized SWCNT samples and highlight the demand of powerful analytical techniques for their investigation. Nevertheless, for an exact determination of the amount of covalently attached addends a reliable calibration for the TG-GC/MS based analysis is a fundamental prerequisite. For this purpose, we systematically varied the amount of $T_{(1:4)}A$ and $G_{(1:4)}A$ (Fig. S20, S21). As the detachment of the addends takes place mainly in the temperature regime between 130 and 500 °C, the detected overall mass loss in this region was weighed by the percentage of peak area of the corresponding *t*-butylbenzene detached from the respective SCA (0.95 for $T_{(1:4)}A$ and 0.16 for $G_{(1:4)}A$) (Fig. S21). For each TG experiment the corresponding amounts of *t*-butylbenzene are extracted (Fig. S23) and were correlated to the peak area of the GC-MS elugram using different weight portions of samples during the thermogravimetric analysis of the specified samples *vs.* the integrated peak area of the GC-MS trace. The linear fit (Figure 6) yields the pre-factor $3.24*10^{-11} \pm 0.14*10^{-11}$ (standard deviation: 4 %) and provides the basis for the direct quantification of the functional groups attached to carbon nanotubes as well as to graphene/graphite. On the basis of the experimentally determined calibration factor, the amount of *t*-butylbenzene in the mixed functionalized samples **T$_{(1:4)}$B** and **G$_{(1:4)}$B** can be calculated as 1.98 µmol and 0.01 µmol, respectively. According to the detected 1:1 ratio of the peak areas (*t*-butylbenzene : bromobenzene) it is furthermore possible to quantify the amount of the detached bromobenzene groups in **T$_{(1:4)}$B** (2.03 µmol) and **T$_{(1:4)}$C** (3.25 µmol) (Fig. S16). An amount of 0.010 µmol was found for **G$_{(1:4)}$B** and 0.026 µmol for **G$_{(1:4)}$C** (Fig. S17).

Furthermore, the amount of potassium used in the SCA activation step (Fig. S24, S25) can be correlated with the final amount of attached functional moieties. The respective data is collected in Table ST2.

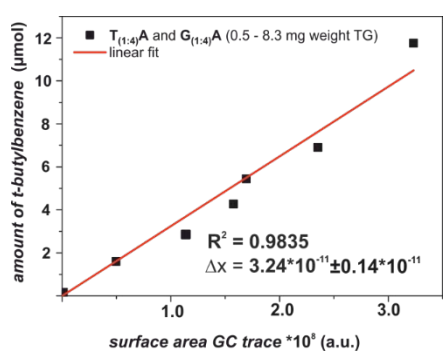

**Figure 6.** TG-GC/MS calibration curve using different sample weights of **T$_{(1:4)}$A** and **G$_{(1:4)}$B** during the TG analysis.

For the first time it is now possible to exactly quantify the degree of functionalization (*f*-addends in relation to SCA lattice carbon atoms - Table 2) of a given functionalization sequence, as the region of mass-loss attributed to the functional group detachment is identified by TG-MS and the amount of a specific covalently bound addend is determined by TG-GC/MS. This is the advantage of this setup as contributions of side products, solvent residues and sample impurities can be identified and the detected sample mass loss, used for the calculation of the degree of functionalization, can exactly be assigned to the specific functional entity of interest.

**Table T2.** Correlation of the amount of attached functional addends with the amount of lattice carbon in the respective functionalized SCA derivatives.

| SWCNT | *f*-addend per lattice carbon / % | graphite graphene | *f*-addend per lattice carbon / % |
|---|---|---|---|
| **T$_{(1:4)}$A** | 3.42 | **G$_{(1:4)}$A** | 0.04 |
| **T$_{(1:4)}$B** | 2.51 | **G$_{(1:4)}$B** | 0.02 |
| **T$_{(1:4)}$C** | 2.02 | **G$_{(1:4)}$C** | 0.01 |
| **T$_{(1:8)}$A** | 2.45 | **G$_{(1:8)}$A** | 0.13 |
| **T$_{(1:16)}$A** | 0.80 | **G$_{(1:16)}$A** | 0.04 |
| **T$_{(1:24)}$A** | 0.25 | **G$_{(1:24)}$A** | 0.02 |

Apparently, the TG-GC/MS determined degrees of addition in the respective SCA derivatives (Fig. S24-S25) nicely correlate with the development of the $I_{(D/G)}$ values, discussed above. To demonstrate that the reductive activation with potassium is indeed crucial for the covalent functionalization with $\lambda^3$-iodane, reference experiments with pristine carbon nanotubes and graphite starting materials have been carried out under various experimental conditions: a) stirring at RT (4 h), b) stirring under light irradiation (4 h) and c) stirring under reflux (4 h). The respective samples have been investigated by Raman spectroscopy and TG-MS (S26-S29). In contrast to the measured $I_{(D/G)}$ ratios for the reductively activated SCAs (0.95 (**T$_{(1:4)}$A**), 1.00 (**G$_{(1:4)}$A**)) only a negligible increase of the $I_{(D/G)}$ (Fig. S26, S28) in relation to the pristine starting allotrope was detected for all three reference experiments. These findings are also corroborated by the corresponding TG-MS data (Fig. S27, S29).

These results nicely illustrate the important role of reductively activated SCA intermediates for the covalent functionalization of carbon allotropes. Together with a suitable electrophile - here $\lambda^3$-iodanes – functional entities can very efficiently be grafted onto the SCA framework. Dyke et al.[12a] as well as Chattopadhyay et al.[12b] have shown, that these types of reductive functionalization sequences are based on single electron transfer processes from the charged carbon allotrope intermediates towards the trapping electrophile. This leads directly to the formation of highly reactive radical species in close proximity to the SCA carbon framework and subsequently, the trapping electrophile is covalently bound. As mentioned above, hypervalent iodine compounds can easily form radicals and therefore a similar radical mechanism can be anticipated for our novel $\lambda^3$-iodane based functionalization route. First evidence for this assumption was obtained from a GC/MS based by-product analysis. Here, the organic components in the solvent phase, obtained after the aqueous work-up of the primary reaction mixture - $\lambda^3$-iodane **C**, were separated by GC and identified by mass spectrometry. Besides 1-bromo-4-iodobenzene as main component, biphenyl derivatives as characteristic radical recombination products were detected (Fig. S30).

**Conclusions**

In conclusion, we have introduced a new $\lambda^3$-iodane based reaction sequence as a versatile and efficient tool for the functionalization of reductively exfoliated and activated SCAs. The covalent aryl binding can be confirmed by SRS and SRM as well as by TG-MS studies. By the aid of temperature dependent Raman investigations it can be shown that the introduced functional moieties can reversibly be detached in a temperature region up to 480 °C. This information provided the basis for TG-GC/MS coupling experiments where it was shown that the thermally detached entities can be identified by MS as one major component after GC separation. These findings enabled us to quantify the amount of covalently bound aryl functionalities, allowing to exactly calculate the percentage of functionalized carbon for each individual component.

**Experimental Section**

**Materials:** Purified HiPco SWCNTs (grade: pure; lot number P0261, TGA residue 13.3 % wt) were purchased from Unidym Inc. (Sunnyvale, CA) and were used without further treatment. Synthetic spherical graphite (SGN18, 99.99 % C, TGA residue 0.01 % wt - Future Carbon, Germany) with a mean grain size of 18 µm and a specific surface area of 6.2 m2/g was used after annealing under vacuum (300 °C).

Chemicals and solvents were purchased from Sigma Aldrich Co. (Germany) and were used as-received if not stated otherwise.

THF was distilled three times in an argon inert gas atmosphere in order to remove residual water: a) over $CaH_2$, b) over sodium and c) over sodium-potassium alloy. Residual traces of oxygen were removed by pump freeze treatment (3 iterative steps). $THF_{(abs)}$ was used for all reactions.

**Glove Box:** Sample functionalization was carried out in an argon filled Labmaster sp glove box (MBraun), equipped with a gas filter to remove solvents and an argon cooling systems, with an oxygen content <0.1 ppm and a water content <0.1 ppm.

**Raman Spectroscopy:** Raman spectroscopic characterization was carried out on a HoribaLabRAM Aramis confocal Raman microscope ($\lambda_{exc}$ : 532, 633, 785 nm) with a laser spot size of about 1 µm (Olympus LMPlanFl 100x, NA 0.80). The incident laser power was kept as low as possible to avoid structural sample damage: 127 µW (532 nm) for SWCNT samples or 1.35 mW (532 nm) for graphite samples, 36 µW (633 nm) and 100 µW (785 nm). Spectra were obtained with a CCD array at -70 °C – grating: 600 grooves/mm [532 & 633 nm] and a 300 groove/mm [785 nm]. Spectra were obtained from a 50 µm x 50 µm area for SWCNTs or 100 x 100 µm area for graphite with 2 µm step size in SWIFT mode for low integration times. Sample movement was carried out by an automated XY-scanning table.

Temperature depending Raman measurements were performed in a Linkam stage THMS 600, equipped with a liquid nitrogen pump TMS94 for temperature stabilization under a constant flow of nitrogen. The measurements were carried out on $Si/SiO_2$ substrates (300 nm oxide layer) with a heating rate of 10 K/min.

**Thermogravimetric Analysis (TG) combined with gaschromaographic separation (GC) coupled with a mass spectrometer (MS):** The thermogravimetric analysis was carried out

on a Perkin Elmer Pyris 1 TGA instrument. Time-dependent temperature profiles in the range of 20 and 700 °C (20 K/min gradient) were carried out under a constant flow of $N_2$ (70 mL/min). About 2.0 mg initial sample mass was used if not stated otherwise. The evolved gases detached from the respective sample in combination with the $N_2$ carrier gas is transferred into the GC system via a TL9000 TG-IR-GC interface at a constant temperature of 280 °C. The gaschromatographic separation was achieved by a GC-Clarus 680 with a polysiloxane coated Elite-5MS capillary column: 30 m length, 0.25 mm diameter, 0.25 μm film thickness. A GC injection fraction of 1 mL was collected at the respective TG temperature: SWCNT samples (250 °C), Graphite samples (300 °C).GC parameters: Injector zone: 280 °C; detection zone: 250 °C; split: 8.2; flow rate helium: 10 mL/min; temperature profile: 34 min total run time; dynamic ramp: 24 min; 40 - 280 °C with a10 K/min gradient followed by an isothermal step of 10 min at 280 °C.

Online MS measurements without GC-separation were carried out with a GC-Clarus 680 with an Elite-5MS glass capillary column: 30 m length, 0.25 mm diameter. GC parameters: Injector zone: 280 °C; detection zone: 250 °C; split: 22.6; flow rate helium: 10 mL/min; isothermal temperature profile: 34 min at 280 °C. MS measurements were performed on a MS Clarus SQ8C (Multiplier: 1800 V). The obtained data was processed with the TurboMass Software and Bibliograpic searches where performed with NIST MS Search 2.0.

**$^1$H, $^{13}$C and $^{19}$F NMR spectroscopy:** NMR spectra were recorded on a Jeol JNM EX 400 (400 MHz for $^1$H and 100 MHz for $^{13}$C) and a Bruker Avance 300 (300 MHz for $^1$H, 75 MHz for $^{13}$C and 282 MHz for $^{19}$F) spectrometer. Chemical shifts are reported in ppm at room temperature (RT). Abbreviations used for splitting patterns are: s = singlet, d = doublet, t = triplet, m = multiplet.

**Synthesis of the $\lambda^3$-Iodanes:** The syntheses of the $\lambda^3$-iodanes were adapted from the work of B. Olofsson and her research group.[13a] *m*-Chloroperbenzoic acid was dried under reduced pressure before use. Dichloromethane was distilled once under reduced pressure before use.

**Synthesis of *f*-SCA:**

Carbon nanotubide salts with varying potassium/carbon ratios:

**T$_{(1:4)}$** [K:C – 1:4 ]; **T$_{(1:8)}$** [K:C – 1:8 ], **T$_{(1:16)}$** [K:C – 1:16], **T$_{(1:24)}$** [K:C – 1:24 ]

In an argon filled glove box (< 0.1 ppm oxygen; <0.1 ppm $H_2O$), 12.00 mg (1.000 mmol) HiPco SWCNTs and the respective amount of potassium – **T$_{(1:4)}$** 9.775 mg [K:C – 1:4 ] (0.250 mmol), **T$_{(1:8)}$** 4.887 mg [K:C – 1:8 ] (0.125 mmol), **T$_{(1:16)}$** 2.444 mg [K:C – 1:16] (0.063 mmol), **T$_{(1:24)}$** 1.667 mg [K:C – 1:24 ] (0.042 mmol) – were heated under stirring at 150 °C for 8 h. Afterwards, the respective salt was allowed to cool to RT and isolated as a black [K:C = 1:24, 1:16, 1:8] or beige [K:C = 1:4] material.

Graphite intercalation compounds with varying potassium/carbon ratios:
**G$_{(1:4)}$** [K:C – 1:4 ]; **G$_{(1:8)}$** [K:C – 1:8 ], **G$_{(1:16)}$** [K:C – 1:16], **G$_{(1:24)}$** [K:C – 1:24 ]

In an argon filled glove box (< 0.1 ppm oxygen; <0.1 ppm H$_2$O), 12.00 mg (1.000 mmol) spherical graphite (SGN18) and the respective amount of potassium – **G$_{(1:4)}$** 9.775 mg [K:C – 1:4 ] (0.250 mmol), **G$_{(1:8)}$** 4.887 mg [K:C – 1:8 ] (0.125 mmol), **G$_{(1:16)}$** 2.444 mg [K:C – 1:16] (0.063 mmol), **G$_{(1:24)}$** 1.667 mg [K:C – 1:24 ] (0.042 mmol) – were heated under stirring at 150 °C for 8 hours. Afterwards, the respective salt was allowed to cool to RT and isolated as a black [K:C = 1:24], bronze [K:C = 1:16, 1:8] or beige [K:C = 1:4] material.

SWCNT: λ$^3$-Iodane variation – preparation of **T$_{(1:4)}$A**, **T$_{(1:4)}$B**, **T$_{(1:4)}$C**

In an argon filled glove box (< 0.1 ppm oxygen; <0.1 ppm H$_2$O), the carbon nanotubide salt **T$_{(1:4)}$**, 21.7 mg salt (1 mmol carbon) and the corresponding λ$^3$-iodane – 293 mg **A** (0.50 mmol), 265 mg **B** (0.50 mmol) and 238 mg **C** (0.50 mmol) – were dispersed by the addition of 20 mL THF$_{(abs)}$ (three different 50 mL round bottom flasks) and by the aid of a 5 min tip ultrasonication treatment step (Bandelin UW 3200, 600 J/min). The respective dispersion (**T$_{(1:4)}$A**, **T$_{(1:4)}$B**, **T$_{(1:4)}$C**) was stirred for 14 h. Afterwards, the reaction mixture was transferred from the glove box and 50 mL of water and a few drops of HCl (until pH = 4 is reached) were added to the dispersion. The reaction mixture was transferred to a separation funnel with 50 mL of cyclohexane. The phases were separated and the organic layer, containing the functionalized SWCNT material, was purged three times with distilled water. The organic layer was filtered through a 0.2 µm reinforced cellulose membrane filter (Sartorius) and washed three times with 100 mL of THF. The covalently functionalized SWCNTs were scrapped off the filter paper and the material was dried in vacuum.

Graphene: λ$^3$-Iodane variation – preparation of **G$_{(1:4)}$A**, **G$_{(1:4)}$B**, **G$_{(1:4)}$C**

In an argon filled glove box (< 0.1 ppm oxygen; <0.1 ppm H$_2$O), the graphite intercalation compound **G$_{(1:4)}$**, 21.7 mg (1 mmol carbon) and the corresponding λ$^3$-iodane – 293 mg **A** (0.50 mmol), 265 mg **B** (0.50 mmol) and 238 mg **C** (0.50 mmol) – were dispersed by the addition of 20 mL THF$_{(abs)}$ (three different 50 mL round bottom flasks) and by the aid of a 5 min tip ultrasonication treatment step (Bandelin UW 3200, 600 J/min). The respective dispersion (**G$_{(1:4)}$A**, **G$_{(1:4)}$B**, **G$_{(1:4)}$C**) was stirred for 14 h. Afterwards, the reaction mixture was transferred from the glove box and 50 mL of water and a few drops of HCl (until pH = 4 is reached) were added to the dispersion. The reaction mixture was transferred to a separation funnel with 50 ml of cyclohexane. The phases were separated and the organic layer, containing the functionalized material, was purged three times with distilled water. The organic layer was filtered through a 0.2 µm reinforced cellulose membrane filter (Sartorius) and washed three times with 100 mL of THF. The covalently functionalized material was dried in vacuum.

SWCNT: Variation of the potassium/carbon ratio - preparation of **T$_{(1:4)}$A, T$_{(1:8)}$A, T$_{(1:16)}$A, T$_{(1:24)}$A**

In an argon filled glove box (< 0.1 ppm oxygen; <0.1 ppm H$_2$O), the respective carbon nanotubide salt (1 mmol carbon) – 21.7 mg (**T$_{(1:4)}$**), 16.8 mg (**T$_{(1:8)}$**), 14.4 mg (**T$_{(1:16)}$**), 13.7 mg (**T$_{(1:24)}$**) – and 293 mg of **A** (0.50 mmol) were dispersed by the addition of 20 mL THF$_{(abs)}$ (four different 50 mL round bottom flasks) and by the aid of a 5 min tip ultrasonication treatment step (Bandelin UW 3200, 600 J/min). The respective dispersion (**T$_{(1:4)}$A, T$_{(1:8)}$A, T$_{(1:16)}$A, T$_{(1:24)}$A**) was stirred for 14 h. Afterwards, the reaction mixture was transferred from the glove box and 50 mL of water and a few drops of HCl (until pH = 4 is reached) were added to the dispersion. The reaction mixture was transferred to a separation funnel with 50 ml of cyclohexane. The phases were separated and the organic layer, containing the functionalized SWCNT material, was purged three times with distilled water. The organic layer was filtered through a 0.2 µm reinforced cellulose membrane filter (Sartorius) and washed three times with 100 mL of THF. The covalently functionalized SWCNTs were scrapped off the filter paper and the material was dried in vacuum.

Graphene: Variation of the potassium/carbon ratio - preparation of **G$_{(1:4)}$A, G$_{(1:8)}$A, G$_{(1:16)}$A, G$_{(1:24)}$A**

In an argon filled glove box (< 0.1 ppm oxygen; <0.1 ppm H$_2$O), the respective graphite intercalation compound (1 mmol carbon) – 21.7 mg (**G$_{(1:4)}$**), 16.8 mg (**G$_{(1:8)}$**), 14.4 mg (**G$_{(1:16)}$**), 13.7 mg (**G$_{(1:24)}$**) – and 293 mg **A** (0.50 mmol) were dispersed by the addition of 20 mL THF$_{(abs)}$ (four different 50 mL round bottom flasks) and by the aid of a 5 min tip ultrasonication treatment step (Bandelin UW 3200, 600 J/min). The respective dispersion (**G$_{(1:4)}$A, G$_{(1:8)}$A, G$_{(1:16)}$A, G$_{(1:24)}$A**) was stirred for 14 h. Afterwards, the samples are removed out of the glove box and 50 mL of water and a few drops of HCl (until pH = 4 is reached) were added to the dispersions. The reaction mixture was transferred to a separation funnel with 50 mL of cyclohexane. The phases were separated and the organic layer, containing the functionalized material, was purged three times with distilled water. The organic layer was filtered through a 0.2 µm reinforced cellulose membrane filter (Sartorius) and washed three times with 100 mL of THF. The covalently functionalized material was dried in vacuum.

**Acknowledgements**

The authors thank the Deutsche Forschungsgemeinschaft (DFG - SFB 953, Project A1 "Synthetic Carbon Allotropes") and the European Research Council (ERC; grant agreement n°246622 GRAPHENOCHEM) for financial support. The research leading to these results has received funding from the European Union Seventh Framework Programme under grant agreement n°604391 Graphene Flagship.

# Novel $\lambda^3$-Iodane Based Functionalization of Synthetic Carbon Allotropes (SCAs) – Common Concepts and Quantification of the Degree of Addition

## Electronic supporting information

*Ferdinand Hof, Ricarda A. Schäfer, Cornelius Weiss, Frank Hauke and Andreas Hirsch\**

## 1. Materials

Purified HiPco SWCNTs (grade: pure; lot number P0261, TGA residue 13.3 % wt) were purchased from Unidym Inc. (Sunnyvale, CA) and were used without further treatment. Synthetic spherical graphite (SGN18, 99.99 % C, TGA residue 0.01 % wt - Future Carbon, Germany) with a mean grain size of 18 µm and a specific surface area of 6.2 m$^2$/g was used after annealing under vacuum (300 °C).

Chemicals and solvents were purchased from Sigma Aldrich Co. (Germany) and were used as-received if not stated otherwise.

THF was distilled three times in an argon inert gas atmosphere in order to remove residual water: a) over CaH$_2$, b) over sodium and c) over sodium-potassium alloy. Residual traces of oxygen were removed by pump freeze treatment (3 iterative steps). THF$_{(abs)}$ was used for all reactions.

## 2. Equipment and Characterization

***Glove Box:*** Sample functionalization was carried out in an argon filled Labmaster sp glove box (MBraun), equipped with a gas filter to remove solvents and an argon cooling systems, with an oxygen content <0.1 ppm and a water content <0.1 ppm.

***Raman Spectroscopy:*** Raman spectroscopic characterization was carried out on a HoribaLabRAM Aramis confocal Raman microscope ($\lambda_{exc}$ : 532, 633, 785 nm) with a laser spot size of about 1 µm (Olympus LMPlanFl 100x, NA 0.80). The incident laser power was kept as low as possible to avoid structural sample damage: 127 µW (532 nm) for SWCNT samples or 1.35 mW (532 nm) for graphite samples, 36 µW (633 nm) and 100 µW (785 nm). Spectra were obtained with a CCD array at -70 °C – grating: 600 grooves/mm [532 & 633 nm] and a 300 groove/mm [785 nm]. Spectra were obtained from a 50 µm x 50 µm area for SWCNTs or 100 x 100 µm area for graphite with 2 µm step size in SWIFT mode for low integration times. Sample movement was carried out by an automated XY-scanning table.

Temperature depending Raman measurements were performed in a Linkam stage THMS 600, equipped with a liquid nitrogen pump TMS94 for temperature stabilization under a constant flow of nitrogen. The measurements were carried out on $Si/SiO_2$ substrates (300 nm oxide layer) with a heating rate of 10 K/min.

**Thermogravimetric Analysis (TG) combined with gaschromaographic separation (GC) coupled with a mass spectrometer (MS): TG-GC-MS Analysis**

The thermogravimetric analysis was carried out on a Perkin Elmer Pyris 1 TGA instrument. Time-dependent temperature profiles in the range of 20 and 700 °C (20 K/min gradient) were carried out under a constant flow of $N_2$ (70 mL/min). About 2.0 mg initial sample mass was used if not stated otherwise.

The evolved gases detached from the respective sample in combination with the $N_2$ carrier gas is transferred into the GC system *via* a TL9000 TG-IR-GC interface at a constant temperature of 280 °C.

The gaschromatographic separation was achieved by a GC-Clarus 680 with a polysiloxane coated Elite-5MS capillary column: 30 m length, 0.25 mm diameter, 0.25 µm film thickness.

A GC injection fraction of 1 mL was collected at the respective TG temperature: SWCNT samples (250 °C), Graphite samples (300 °C).

GC parameters: Injector zone: 280 °C; detection zone: 250 °C; split: 8.2; flow rate helium: 10 mL/min; temperature profile: 34 min total run time; dynamic ramp: 24 min; 40 - 280 °C with a10 K/min gradient followed by an isothermal step of 10 min at 280 °C.

Online MS measurements without GC-separation were carried out with a GC-Clarus 680 with an Elite-5MS glass capillary column: 30 m length, 0.25 mm diameter.

GC parameters: Injector zone: 280 °C; detection zone: 250 °C; split: 22.6; flow rate helium: 10 mL/min; isothermal temperature profile: 34 min at 280 °C.

MS measurements were performed on a MS Clarus SQ8C (Multiplier: 1800 V). The obtained data was processed with the TurboMass Software and Bibliograpic searches where performed with NIST MS Search 2.0.

*$^1$H, $^{13}$C and $^{19}$F NMR spectroscopy:* NMR spectra were recorded on a Jeol JNM EX 400 (400 MHz for $^1$H and 100 MHz for $^{13}$C) and a Bruker Avance 300 (300 MHz for $^1$H, 75 MHz for $^{13}$C and 282 MHz for $^{19}$F) spectrometer. Chemical shifts are reported in ppm at room temperature (RT). Abbreviations used for splitting patterns are: s = singlet, d = doublet, t = triplet, m = multiplet.

## 3. Synthesis of the Hexyl-Functionalized SWCNT Reference Sample

Sidewall functionalized SWCNT derivatives have been synthesized according to the procedure published earlier:[1]

In an argon filled glove box (<0.1 ppm oxygen; <0.1 ppm $H_2O$), 24 mg (2.00 mmol) HiPco SWCNTs and 39.1 mg (1.0 mmol) potassium were heated (150 °C) under stirring for 6 h. The formed carbon nanotubide salt was transferred in a flame-dried round bottom flask equipped with a gas inlet. 50 mL of THF$_{(abs)}$ was added, followed by a short ultrasonication step with a tip sonicator (Bandelin UW 3200, 5 min, 100 J/min). 1.49 mg (10 mmol, 5 eq./C) of *n*-hexyl iodide was added to this dispersion and the reaction mixture was stirred for 1 h at RT. Afterwards, the dispersion was removed from the glove box and transferred to a separation funnel with cyclohexane and water (50 mL, each). The water/THF phase was discarded and the cyclohexane layer with the *n*-hexyl functionalized nanotubes was purged three times with water. Afterwards, the organic layer was filtered through a 0.2 µm reinforced cellulose membrane filter (Sartorius) and washed with 100 mL of THF. The functionalized SWCNT derivatives were scraped off the filter paper and the resulting black powder was dried at RT in vacuum.

# 4. Synthesis, Purification and Characterization of the λ³–Iodanes

*General Information*

The syntheses of the λ³-iodanes were adapted from the work of B. Olofsson and her research group.[2] *m*-Chloroperbenzoic acid was dried under reduced pressure before use. Dichloromethane was distilled once under reduced pressure before use.

[2]M. Bielawski, B. Olofsson, *Chem. Commun.* **2007**, 2521-2523

*Synthesis of bis-(4-(t-butyl)phenyl)-iodonium trifluoromethanesulfonate:* **A**

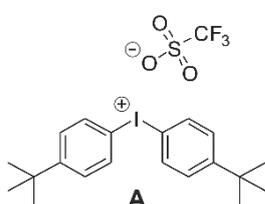

In a 500 mL round bottom flask, 1.93 g (8.50 mmol) iodine was dissolved in 200 mL dichloromethane. 4.39 g (25.5 mmol) of *m*-chloroperbenzoic acid was added to the solution under constant stirring. The reaction mixture was refluxed for 20 min until the oxidation was completed. The reaction mixture was cooled to 0 °C (ice bath) and subsequently 5.27 mL (34.0 mmol) of *t*-butylbenzene was added and the solution was kept stirring for 30 min. 3.01 mL (34.0 mmol) of trifluoromethanesulfonic acid was added dropwise by a Hamilton syringe under violent stirring at 0 °C. The reaction mixture was allowed to warm to RT and was kept stirring for 6 h. Afterwards, 100 mL of water was added to the dispersion. The phases were separated by the aid of a separation funnel and the organic layer was extracted 3 times with 50 mL water. The organic solvent was removed under reduced pressure, yielding an orange solid, which was subsequently dissolved with 50 mL of cold diethyl ether. The solution was stirred until beige-brown crystals started to precipitate. For a complete crystallization, the dispersion was stored at -20 °C for a few hours. Afterwards, the solid was isolated by filtration and was recrystallized from diethyl ether several times, until white needle shaped crystals were yielded. After drying under reduced pressure, λ³-iodane **A** was isolated as white solid. Yield: 2.97 g (5.48 mmol), 64.5 %.

**¹H-NMR** (400 MHz, CDCl₃): 1.27 ppm (s, 18 H); 7.42 ppm (d, *J* = 12.00 Hz, 4 H); 7.86 ppm (d, *J* = 8.05 Hz, 4 H)
**¹³C-NMR** (100 MHz, CDCl₃): 30.95 ppm; 35.21 ppm; 109.45 ppm; 129.60 ppm; 134.84 ppm; 156.56 ppm
**¹⁹F-NMR** (282 MHz, acetone-d6): -75.89 ppm

*Synthesis of (4-bromophenyl)(4-(t-butyl)phenyl)iodonium trifluoromethanesulfonate:* **B**

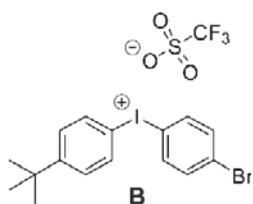

In a 500 mL round bottom flask, 1.65 g 1-bromo-4-iodobenzene (8.85 mmol) was dissolved in 200 mL dichloromethane. 1.52 g (8.85 mmol) of *m*-chloroperbenzoic acid was added to the solution under constant stirring. The reaction mixture was refluxed for 20 min until the oxidation was completed. The reaction mixture was cooled to 0 °C (ice bath) and subsequently 1.19 g (8.85 mmol) of *t*-butylbenzene was added and the solution was kept stirring for 20 min. 3.56 mL (17.7 mmol) of trifluoromethanesulfonic acid was added dropwise by a Hamilton syringe under violent stirring at 0 °C – color change from yellow, over green to black. The reaction mixture was allowed to warm to RT and was kept stirring for 6 h. Afterwards, 100 mL of water was added to the dispersion. The phases were separated by the aid of a separation funnel and the organic layer was extracted 2 times with 50 mL water. The organic solvent was removed under reduced pressure, yielding a black-brown solid, which was subsequently dissolved in a minimum amount of warm methanol. By the addition of 50 mL of cold diethyl ether **S2** started to crystalize. For a complete crystallization, the dispersion was stored at -20 °C for a few hours. Afterwards, the solid was isolated by filtration and was recrystallized from diethyl ether several times, until white needle shaped crystals were yielded. After drying under reduced pressure, $\lambda^3$-iodane **B** was isolated as white solid. Yield: 2.41 g (4.26 mmol), 48.2 %.

**$^1$H-NMR** (400 MHz, CD$_3$OD): 1.31 ppm (s, 9 H); 7.57 ppm (d, *J* = 8.80 Hz, 2 H); 7.68 ppm (d, *J* = 6.80 Hz, 2 H); 8.04 ppm (d, *J* = 8.40 Hz, 2 H); 8.07 ppm (d, *J* = 6.80 Hz, 2 H)
**$^{13}$C-NMR** (100 MHz, CD$_3$OD): 31.31 ppm; 36,16 ppm; 112.69 ppm; 114.25 ppm; 128.54 ppm; 130.62 ppm; 136.23 ppm; 136.27 ppm; 137.95 ppm; 158.09 ppm
**$^{19}$F-NMR** (282 MHz, CD$_3$OD): -76.65 ppm

*Synthesis of bis-(4-bromophenyl)-iodonium trifluoromethanesulfonate:* **C**

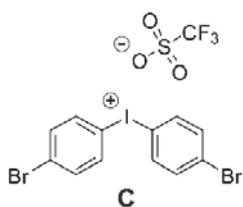

In a 500 mL round bottom flask, 1.93 g iodine (8.50 mmol) was dissolved in 200 mL dichloromethane. 4.39 g *m*-chloroperbenzoic acid (25.5 mmol) was added to the solution under constant stirring. The reaction mixture was refluxed for 20 min until the oxidation was completed. The reaction mixture was cooled to 0 °C (ice bath) and subsequently 3.56 mL bromobenzene (34.0 mmol) was added and the solution was kept stirring for 30 min. 3.01 mL trifluoromethanesulfonic acid (34.0 mmol) was added dropwise by a Hamilton syringe under violent stirring at 0 °C. The reaction mixture was allowed to warm to RT and was kept stirring for 6 h. Afterwards, 100 mL of water was added to the dispersion. The phases were separated by the aid of a separation funnel and the organic layer was extracted 3 times with 50 mL water. The organic solvent was removed under reduced pressure, yielding an orange solid, which was subsequently dissolved with 50 mL of cold diethyl ether. The solution was stirred until beige-brown crystals started to precipitate. For a complete crystallization, the dispersion was stored at -20 °C for a few hours. Afterwards, the solid was isolated by filtration and was recrystallized from diethyl ether several times, until white needle shaped crystals were yielded. After drying under reduced pressure, $\lambda^3$-iodane **C** was isolated as white solid. Yield: 3.51 g (6.05 mmol), 71.2 %.

**[1]H-NMR** (400 MHz, acetone-d6): .77 ppm (d, J = 8.91 Hz, 4 H); 8.28 ppm (d, J = 8.79 Hz, 4 H)
**[13]C-NMR** (100 MHz, acetone-d6): 113.71 ppm; 128.23 ppm; 136.09 ppm; 138.34 ppm
**[19]F-NMR** (282 MHz, acetone-d6): -76.65 ppm

## 5. Synthesis of the Covalently Functionalized Carbon Allotropes

*Carbon nanotubide salts with varying potassium/carbon ratios:*
**T(1:4)** [K:C :$\frac{1}{4}$]; **T(1:8)** [K:C :$\frac{1}{8}$], **T(1:16)** [K:C :$\frac{1}{16}$], **T(1:24)** [K:C :$\frac{1}{24}$]

In an argon filled glove box (< 0.1 ppm oxygen; <0.1 ppm $H_2O$), 12.00 mg (1.000 mmol) HiPco SWCNTs and the respective amount of potassium – **T(1:4)** 9.775 mg [K:C :$\frac{1}{4}$] (0.250 mmol), **T(1:8)** 4.887 mg [K:C :$\frac{1}{8}$] (0.125 mmol), **T(1:16)** 2.444 mg [K:C :$\frac{1}{16}$] (0.063 mmol), **T(1:24)** 1.667 mg [K:C :$\frac{1}{24}$] (0.042 mmol) – were heated under stirring at 150 °C for 8 h. Afterwards, the respective salt was allowed to cool to RT and isolated as a black [K:C = 1:24, 1:16, 1:8] or beige [K:C = 1:4] material.

Graphite intercalation compounds with varying potassium/carbon ratios:
**G(1:4)** [K:C : $\frac{1}{4}$ ]; **G(1:8)** [K:C : $\frac{1}{8}$ ], **G(1:16)** [K:C : $\frac{1}{16}$ ], **G(1:24)** [K:C : $\frac{1}{24}$ ]

In an argon filled glove box (< 0.1 ppm oxygen; <0.1 ppm $H_2O$), 12.00 mg (1.000 mmol) spherical graphite (SGN18) and the respective amount of potassium – **G(1:4)** 9.775 mg [K:C : $\frac{1}{4}$ ] (0.250 mmol), **G(1:8)** 4.887 mg [K:C : $\frac{1}{8}$ ] (0.125 mmol), **G(1:16)** 2.444 mg [K:C : $\frac{1}{16}$ ] (0.063 mmol), **G(1:24)** 1.667 mg [K:C : $\frac{1}{24}$ ] (0.042 mmol) – were heated under stirring at 150 °C for 8 hours. Afterwards, the respective salt was allowed to cool to RT and isolated as a black [K:C = 1:24], bronze [K:C = 1:16, 1:8] or beige [K:C = 1:4] material.

*SWCNT:$\lambda^3$-Iodane variation – preparation of **T(1.4)A**, **T(1,4)B**, **T(1:4)C***

In an argon filled glove box (< 0.1 ppm oxygen; <0.1 ppm $H_2O$), the carbon nanotubide salt **T(1:4)**, 21.7 mg salt (1 mmol carbon) and the corresponding $\lambda^3$-iodane – 293 mg **A** (0.50 mmol), 265 mg **B** (0.50 mmol) and 238 mg **C** (0.50 mmol) – were dispersed by the addition of 20 mL THF$_{(abs)}$ (three different 50 mL round bottom flasks) and by the aid of a 5 min tip ultrasonication treatment step (Bandelin UW 3200, 600 J/min). The respective dispersion (**T(1:4)A**, **T(1:4)B**, **T(1:4)C**) was stirred for 14 h. Afterwards, the reaction mixture was transferred from the glove box and 50 mL of water and a few drops of HCl (until pH = 4 is reached) were added to the dispersion. The reaction mixture was transferred to a separation funnel with 50 mL of cyclohexane. The phases were separated and the organic layer, containing the functionalized SWCNT material, was purged three times with distilled water. The organic layer was filtered through a 0.2 μm reinforced cellulose membrane filter (Sartorius) and washed three times with 100 mL of THF. The covalently functionalized SWCNTs were scrapped off the filter paper and the material was dried in vacuum.

*Graphene: λ³-Iodane variation – preparation of $G_{(1.4)}A$, $G_{(1,4)}B$, $G_{(1:4)}C$*

In an argon filled glove box (< 0.1 ppm oxygen; <0.1 ppm $H_2O$), the graphite intercalation compound $G_{(1,4)}$, 21.7 mg (1 mmol carbon) and the corresponding $λ^3$-iodane – 293 mg **A** (0.50 mmol), 265 mg **B** (0.50 mmol) and 238 mg **C** (0.50 mmol) – were dispersed by the addition of 20 mL $THF_{(abs)}$ (three different 50 mL round bottom flasks) and by the aid of a 5 min tip ultrasonication treatment step (Bandelin UW 3200, 600 J/min) The respective dispersion ($G_{(1:4)}A$, $G_{(1:4)}B$, $G_{(1:4)}C$) was stirred for 14 h. Afterwards, the reaction mixture was transferred from the glove box and 50 mL of water and a few drops of HCl (until pH = 4 is reached) were added to the dispersion. The reaction mixture was transferred to a separation funnel with 50 ml of cyclohexane. The phases were separated and the organic layer, containing the functionalized material, was purged three times with distilled water. The organic layer was filtered through a 0.2 μm reinforced cellulose membrane filter (Sartorius) and washed three times with 100 mL of THF. The covalently functionalized material was dried in vacuum.

*SWCNT: Variation of the potassium/carbon ratio - preparation of $T_{(1:4)}A$, $T_{(1:8)}A$, $T_{(1:16)}A$, $T_{(1:24)}A$*

In an argon filled glove box (< 0.1 ppm oxygen; <0.1 ppm $H_2O$), the respective carbon nanotubide salt (1 mmol carbon) – 21.7 mg ($T_{(1:4)}$), 16.8 mg ($T_{(1:8)}$), 14.4 mg ($T_{(1:16)}$), 13.7 mg ($T_{(1:24)}$) – and 293 mg of **A** (0.50 mmol) were dispersed by the addition of 20 mL $THF_{(abs)}$ (four different 50 mL round bottom flasks) and by the aid of a 5 min tip ultrasonication treatment step (Bandelin UW 3200, 600 J/min) The respective dispersion ($T_{(1:4)}A$, $T_{(1:8)}A$, $T_{(1:16)}A$, $T_{(1:24)}A$) was stirred for 14 h. Afterwards, the reaction mixture was transferred from the glove box and 50 mL of water and a few drops of HCl (until pH = 4 is reached) were added to the dispersion. The reaction mixture was transferred to a separation funnel with 50 ml of cyclohexane. The phases were separated and the organic layer, containing the functionalized SWCNT material, was purged three times with distilled water. The organic layer was filtered through a 0.2 μm reinforced cellulose membrane filter (Sartorius) and washed three times with 100 mL of THF. The covalently functionalized SWCNTs were scrapped off the filter paper and the material was dried in vacuum.

*Graphene: Variation of the potassium/carbon ratio - preparation of $G_{(1:4)}A$, $G_{(1:8)}A$, $G_{(1:16)}A$, $G_{(1:24)}A$*

In an argon filled glove box (< 0.1 ppm oxygen; <0.1 ppm H$_2$O), the respective graphite intercalation compound (1 mmol carbon) – 21.7 mg (**G$_{(1:4)}$**), 16.8 mg (**G$_{(1:8)}$**), 14.4 mg (**G$_{(1:16)}$**), 13.7 mg (**G$_{(1:24)}$**) – and 293 mg **A** (0.50 mmol) were dispersed by the addition of 20 mL THF$_{(abs)}$ (four different 50 mL round bottom flasks) and by the aid of a 5 min tip ultrasonication treatment step (Bandelin UW 3200, 600 J/min) The respective dispersion (**G$_{(1:4)}$A**, **G$_{(1:8)}$A**, **G$_{(1:16)}$A**, **G$_{(1:24)}$A**) was stirred for 14 h. Afterwards, the samples are removed out of the glove box and 50 mL of water and a few drops of HCl (until pH = 4 is reached) were added to the dispersions. The reaction mixture was transferred to a separation funnel with 50 mL of cyclohexane. The phases were separated and the organic layer, containing the functionalized material, was purged three times with distilled water. The organic layer was filtered through a 0.2 µm reinforced cellulose membrane filter (Sartorius) and washed three times with 100 mL of THF. The covalently functionalized material was dried in vacuum.

## 6. Reference Experiments with Pristine Starting Materials

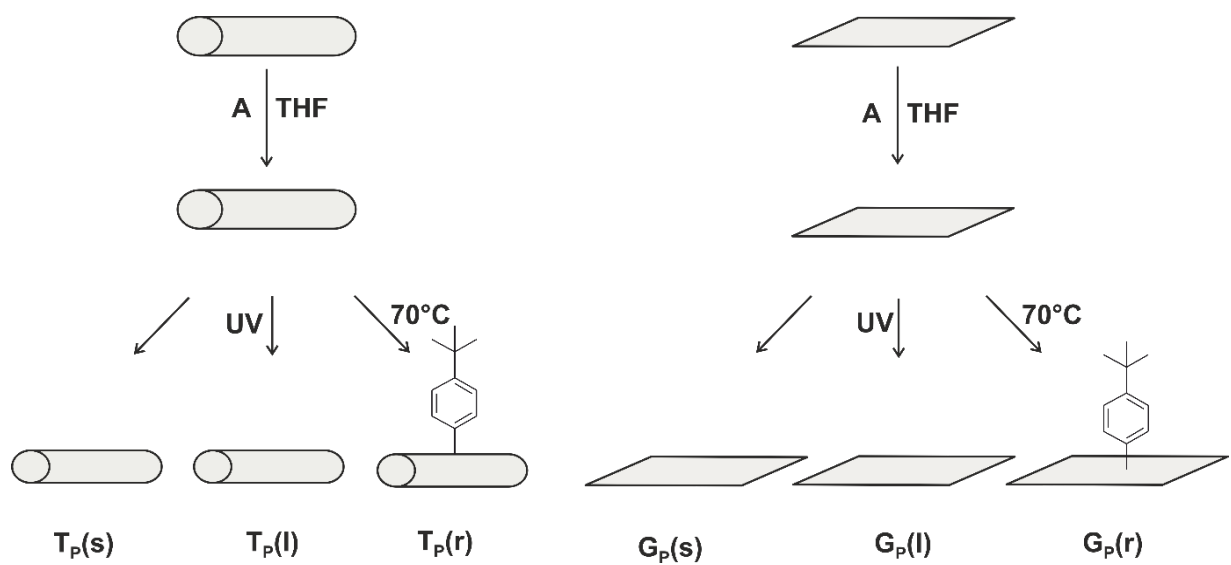

**Scheme S1:** Reference experiments with the two carbon allotropes (left: SWCNTs, right: graphene) in the presence of λ-iodane **A** under different reaction conditions: a) stirring, b) UV light irradiation and c) reflux at 70 °C.

*SWCNT: Pristine SWCNTs with $\lambda^3$-iodane under varying reaction conditions – stirring (s), light irradiation (l) and reflux (r)*

In three 100 mL flame-dried, one-neck Schlenk flasks, 12 mg (1 mmol) HiPco SWCNTs and 293 mg of **A** (0.5 mmol) were dispersed in THF$_{(abs)}$ (20 mL, each) by the aid of a 5 min tip ultrasonication treatment step (Bandelin UW 3200, 600 J/min). One dispersion was stirred for 4 h (yielding **T$_P$(s)**), one was stirred for 4 h under UV-light irritation (yielding **T$_P$(l)**) and one was stirred for 4 h under reflux at 70 °C (yielding **T$_P$(r)**). Afterwards, 50 mL of water and a few drops HCl (until pH =4 is reached) was added to the respective reaction mixture and each dispersion was transferred to a separation funnel with 50 mL of cyclohexane. The phases were separated and the organic layer, containing the functionalized SWCNT material, was purged three times with distilled water. Afterwards, the organic layer was filtered through a 0.2 µm reinforced cellulose membrane filter (Sartorius) and washed three times with 100 mL of THF. The material was scrapped off the filter paper and the material was dried in vacuum.

*Graphene: Pristine graphite (SGN18) with $\lambda^3$-iodane under varying reaction conditions – stirring (s), light irradiation (l) and reflux (r)*

In three 100 mL flame-dried, one-neck Schlenk flasks, 12 mg (1 mmol) spherical graphite (SGN18) and 293 mg **A** (0.5 mmol) were dispersed in THF$_{(abs)}$ (20 mL, each) by the aid of a 5 min tip ultrasonication treatment step (Bandelin UW 3200, 600 J/min).. One dispersion was stirred for 4 h (yielding **G$_P$(s)**), one was stirred for 4 h under UV-light irritation (yielding **G$_P$(l)**) and one was stirred for 4 h under reflux at 70 °C (yielding **G$_P$(r)**) Afterwards, 50 mL of water and a few drops HCl (until pH =4 is reached) was added to the respective reaction mixture and each dispersion was transferred to a separation funnel with 50 mL of cyclohexane. The phases were separated and the organic layer, containing the functionalized material, was purged three times with distilled water. Afterwards, the organic layer was filtered through a 0.2 µm reinforced cellulose membrane filter (Sartorius) and washed three times with 100 mL of THF. The material was scrapped off the filter paper and the material was dried in vacuum.

## 7. Additional Raman Data

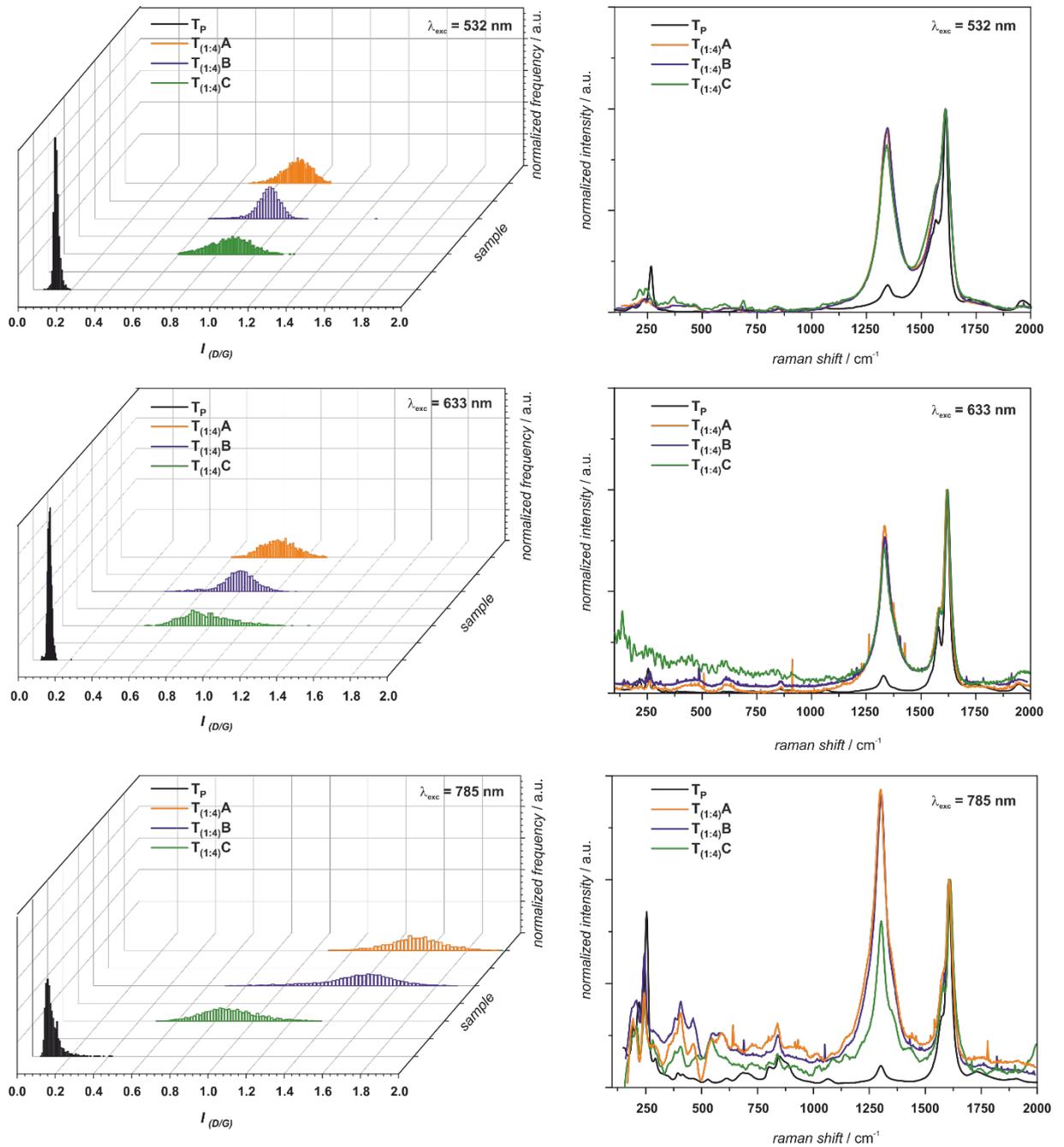

**Figure S1:** Scanning Raman Spectroscopy (SRS) of the samples **T$_{(1:4)}$A** (orange trace), **T$_{(1:4)}$B** (blue trace), and **T$_{(1:4)}$C** (green trace) carried out at different laser excitation wavelengths – top: $\lambda_{exc}$ = 532 nm, middle: $\lambda_{exc}$ = 633 nm, bottom: $\lambda_{exc}$ = 785 nm. Left column: Corresponding histograms of the respective I$_{(D/G)}$ values. Right column: Mean spectra of the respective SWCNT sample.

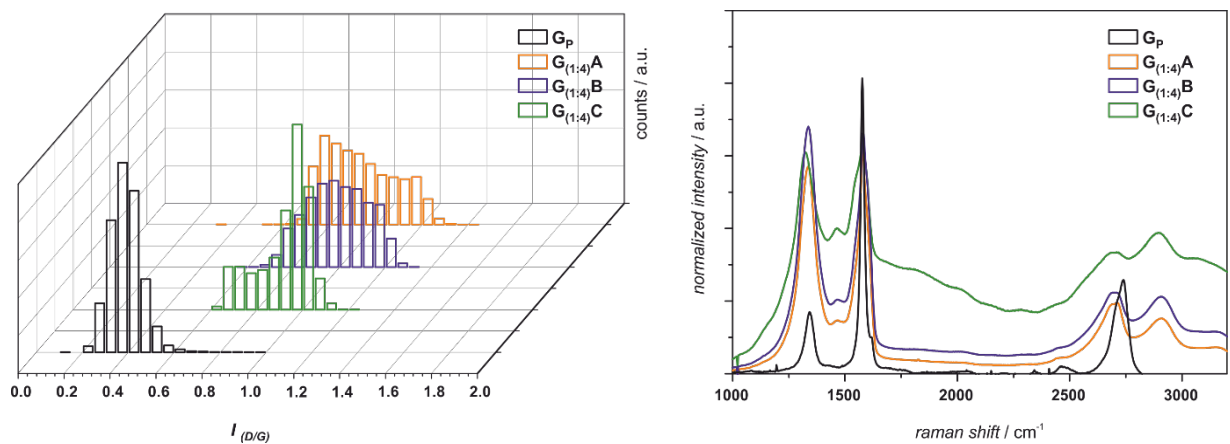

**Figure S2:** Scanning Raman Spectroscopy (SRS) of the samples **G$_{(1:4)}$A** (orange trace), **G$_{1(1:4)}$B** (blue trace), and **G$_{(1:4)}$C** (green trace) carried out at $\lambda_{exc}$ = 532 nm excitation wavelength. Left: Corresponding histograms of the respective I$_{(D/G)}$ values. Right: Mean spectra of the respective graphene/graphite sample.

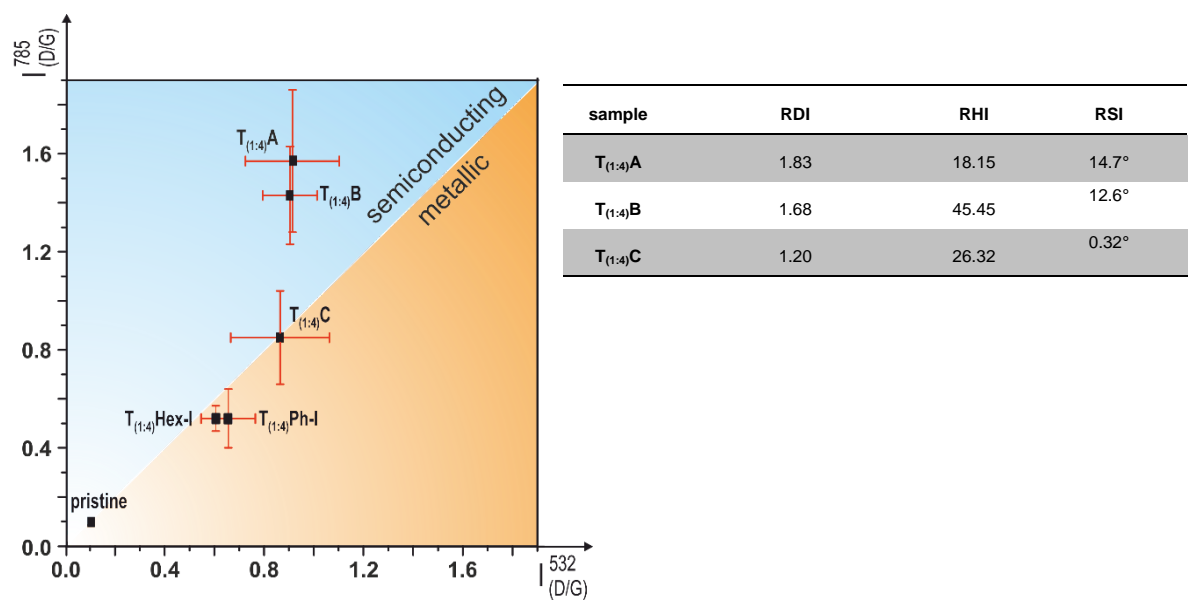

| sample | RDI | RHI | RSI |
|---|---|---|---|
| T(1:4)A | 1.83 | 18.15 | 14.7° |
| T(1:4)B | 1.68 | 45.45 | 12.6° |
| T(1:4)C | 1.20 | 26.32 | 0.32° |

**Figure S3:** Left: 2D Raman index plot based on the statistical Raman analysis of **T(1:4)A**, **T(1:4)B** and **T(1:4)C**. Also shown: results for the alkylated and arylated reference samples **T(1:4)Hex-I** and **T(1:4)Ph-I**. Right: The respective values for the Raman Defect Index (RDI), Raman Homogeneity Index (RHI), and Raman.[3]

[3]F. Hof, S. Bosch, J. M. Englert, F. Hauke and A. Hirsch, *Angew. Chem. Int. Ed.*, 2012, **51**, 11727

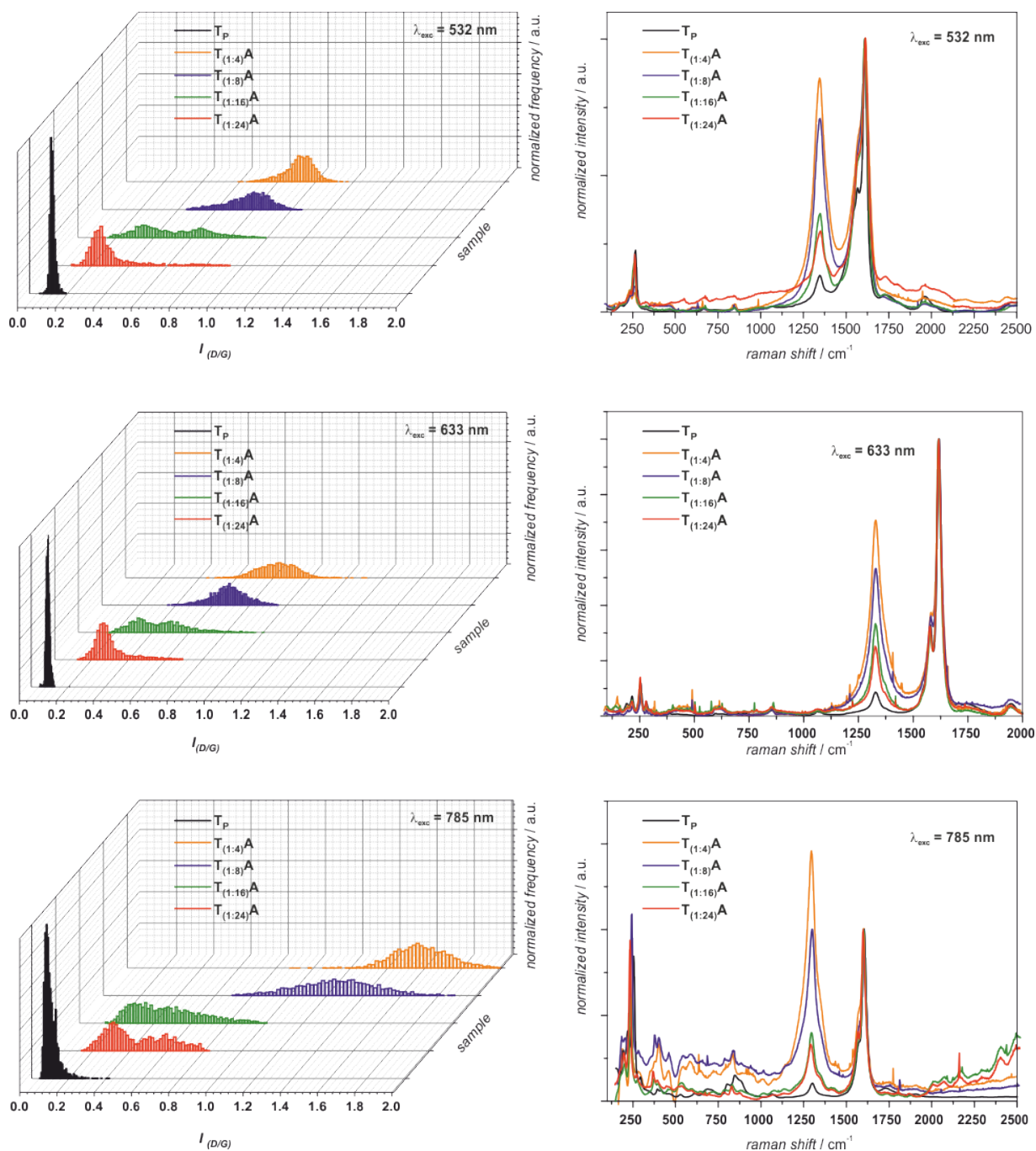

**Figure S4:** Scanning Raman Spectroscopy (SRS) of the samples **T$_{(1:4)}$A** (orange trace), **T$_{(1:8)}$A** (blue trace), **T$_{(1:16)}$A** (green trace), and **T$_{(1:24)}$A** (red trace) carried out at different laser excitation wavelengths – top: $\lambda_{exc}$ = 532 nm, middle: $\lambda_{exc}$ = 633 nm, bottom: $\lambda_{exc}$ = 785 nm. Left column: Corresponding histograms of the respective I$_{(D/G)}$ values. Right column: Mean spectra of the respective SWCNT sample.

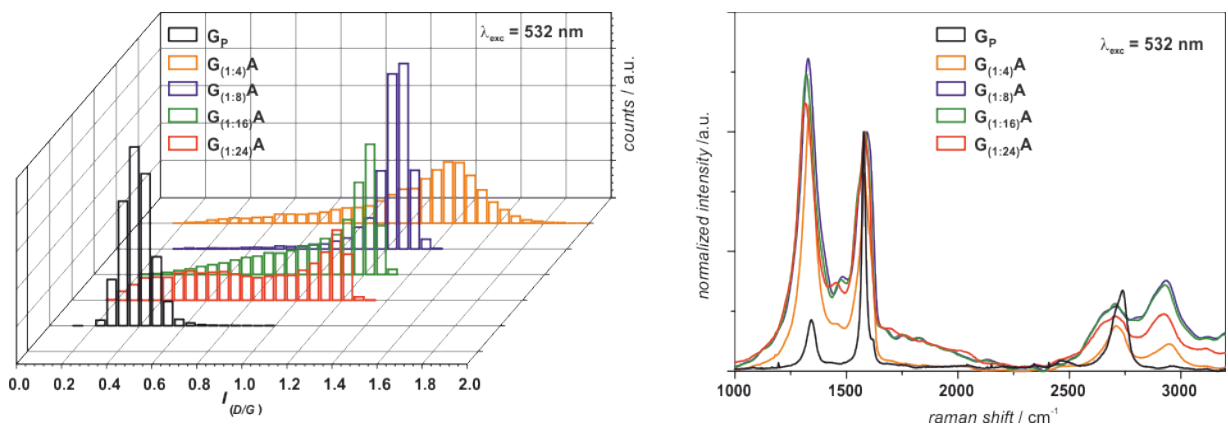

**Figure S5:** Scanning Raman Spectroscopy (SRS) of the samples **G$_{(1:4)}$A** (orange trace), **G$_{(1:8)}$A** (blue trace), **G$_{(1:16)}$A** (green trace), and **G$_{(1:24)}$A** (red trace) carried out at $\lambda_{exc}$ = 532 nm excitation wavelength. Left: Corresponding histograms of the respective I$_{(D/G)}$ values. Right: Mean spectra of the respective SWCNT sample.

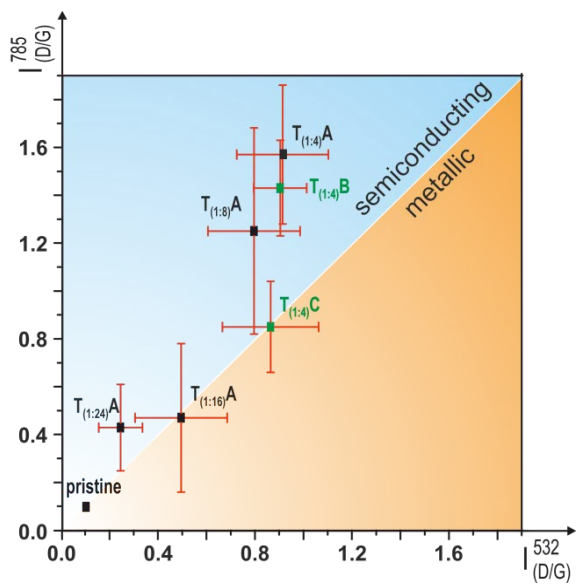

**Figure S6:** 2D Raman index plot based on the statistical Raman analysis of **T$_{(1:4)}$A**, **T$_{(1:4)}$B** and **T$_{(1:4)}$C** varying the $\lambda^3$-iodane during the synthesis in comparison to the samples **T$_{(1:8)}$A**, **T$_{(1:16)}$A**, **T$_{(1:24)}$A** varying the potassium concentration in the activation step.[3]

[3] F. Hof, S. Bosch, J. M. Englert, F. Hauke and A. Hirsch, *Angew. Chem. Int. Ed.*, 2012, **51**, 11727

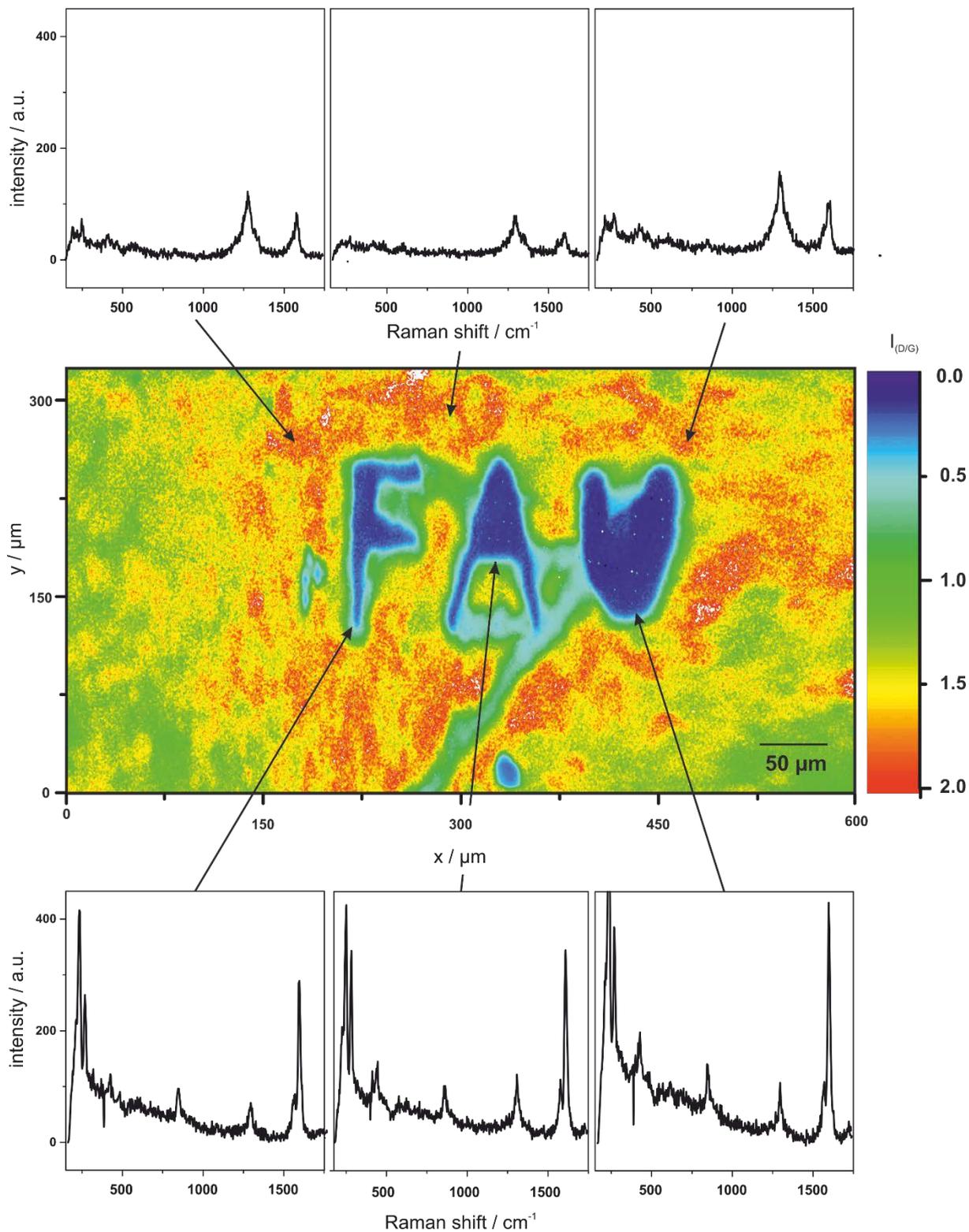

**Figure S7:** Laser induced thermal defunctionalization of **T$_{(1:4)}$A** with corresponding Raman spectra of the different regions of the bulk sample. Excitation wavelength: 785 nm.

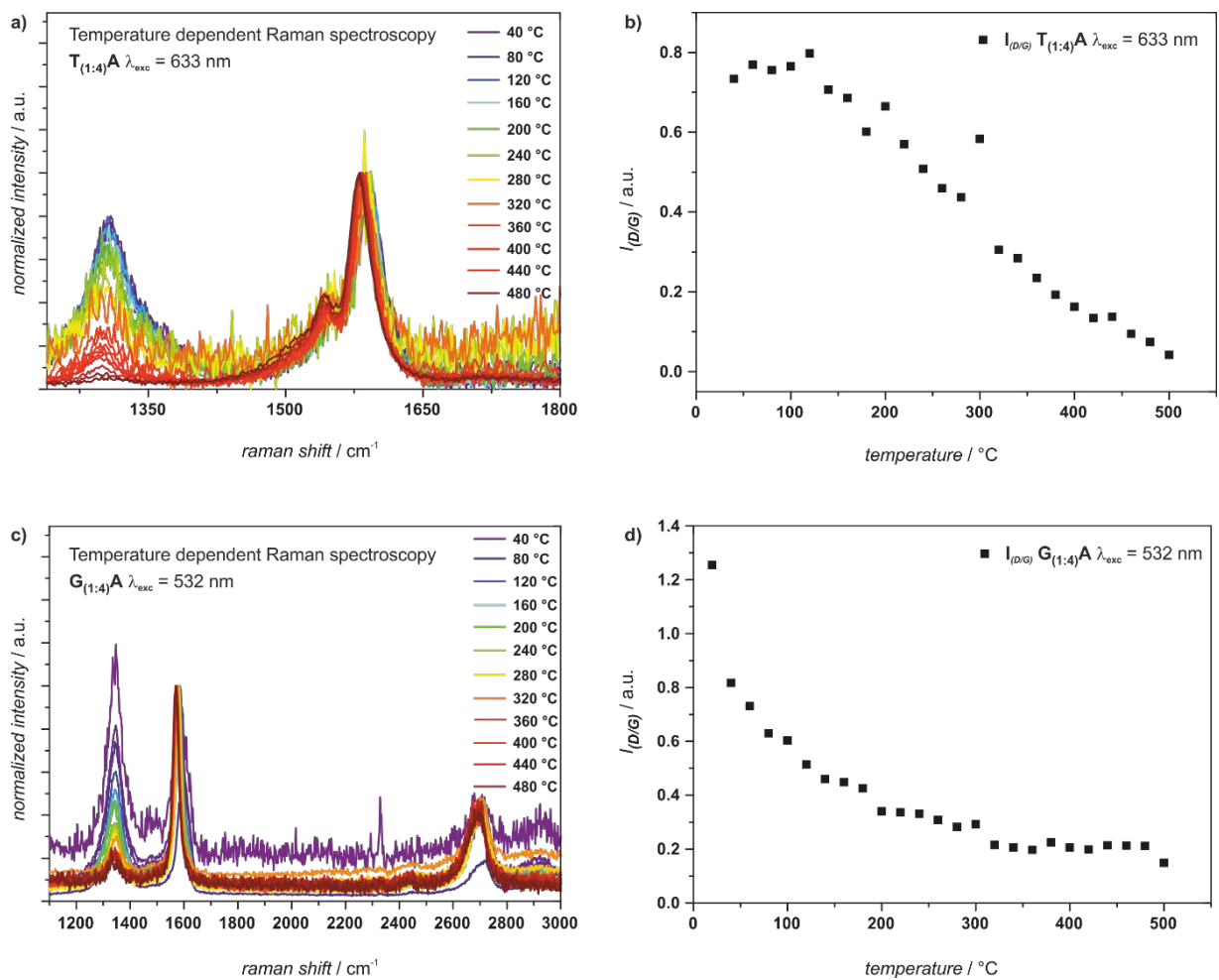

**Figure S8:** Temperature dependent Raman spectroscopy of *t*-butylphenyl functionalized carbon allotropes. a) Raman spectra of the carbon nanotube sample **T$_{(1:4)}$A** in the temperature range of 20-500 °C, $\lambda_{exc}$ = 633 nm, $\Delta$T = 20 K steps. b) Extracted I$_{(D/G)}$ values *vs.* temperature. c) Measurement of the graphite sample **G$_{(1:4)}$A** in the temperature range of 20-500 °C, $\lambda_{exc}$ = 633 nm, $\Delta$T = 20 K steps. d) Extracted I$_{(D/G)}$ values *vs.* temperature.

## 8. Additional TG-MS Data

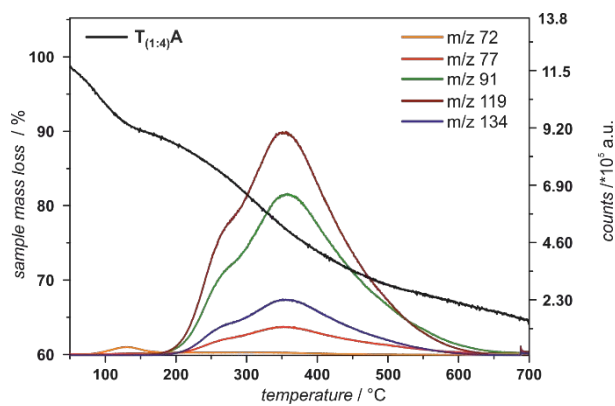

**Figure S9:** TG-MS analysis of the *t*-butylphenyl functionalized carbon nanotube sample **T$_{(1:4)}$A**. Depicted MS traces can be attributed to *t*-butylphenyl fragments – m/z (77, 91, 119 and 134) – and the solvent THF m/z (72).

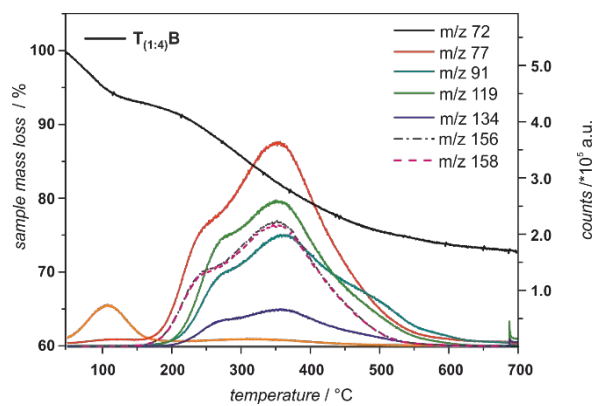

**Figure S10:** TG-MS analysis of mixed functionalized carbon nanotube sample **T$_{(1:4)}$B**. Depicted MS traces can be attributed to *t*-butylphenyl fragments – m/z (77, 91, 119 and 134) – and bromobenzene fragments – m/z (77, 156, 158) – and the solvent THF m/z (72).

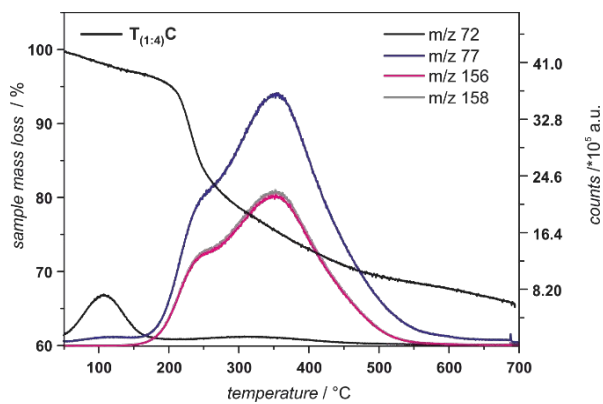

**Figure S11:** TG-MS analysis of the bromobenzene functionalized carbon nanotube sample **T$_{(1:4)}$C**. Depicted MS traces can be attributed to bromobenzene fragments – m/z (77, 156 and 158) – and the solvent THF m/z (72).

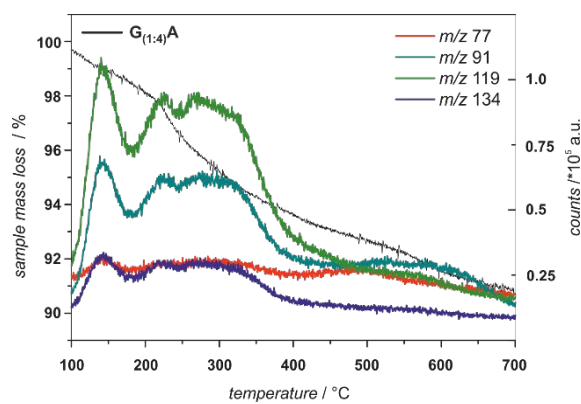

**Figure S12:** TG-MS analysis of the *t*-butylphenyl functionalized graphite sample **G$_{(1:4)}$A**. Depicted MS traces can be attributed to *t*-butylphenyl fragments – m/z (77, 91, 119 and 134) – and the solvent THF m/z (72).

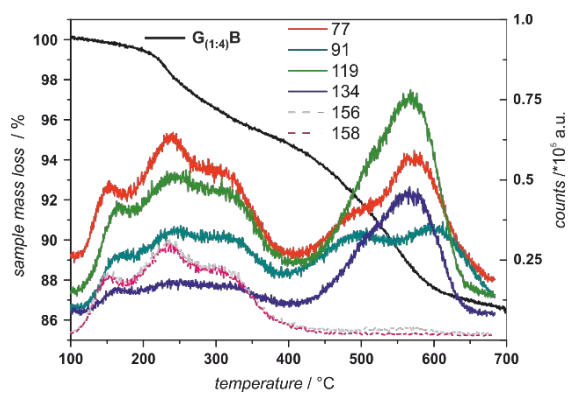

**Figure S13:** TG-MS analysis mixed functionalized graphite sample **G$_{(1:4)}$B**. Depicted MS traces can be attributed to *t*-butylphenyl fragments – m/z (77, 91, 119 and 134) – bromobenzene fragments – m/z (77, 156, 158) – and the solvent THF m/z (72).

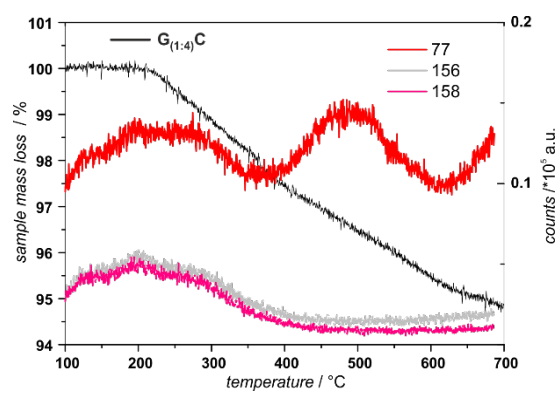

**Figure S14:** TG-MS analysis of the bromobenzene functionalized graphite sample **G$_{(1:4)}$C**. Depicted MS traces can be attributed to fragments – m/z (77, 156 and 158) – and the solvent THF m/z (72).

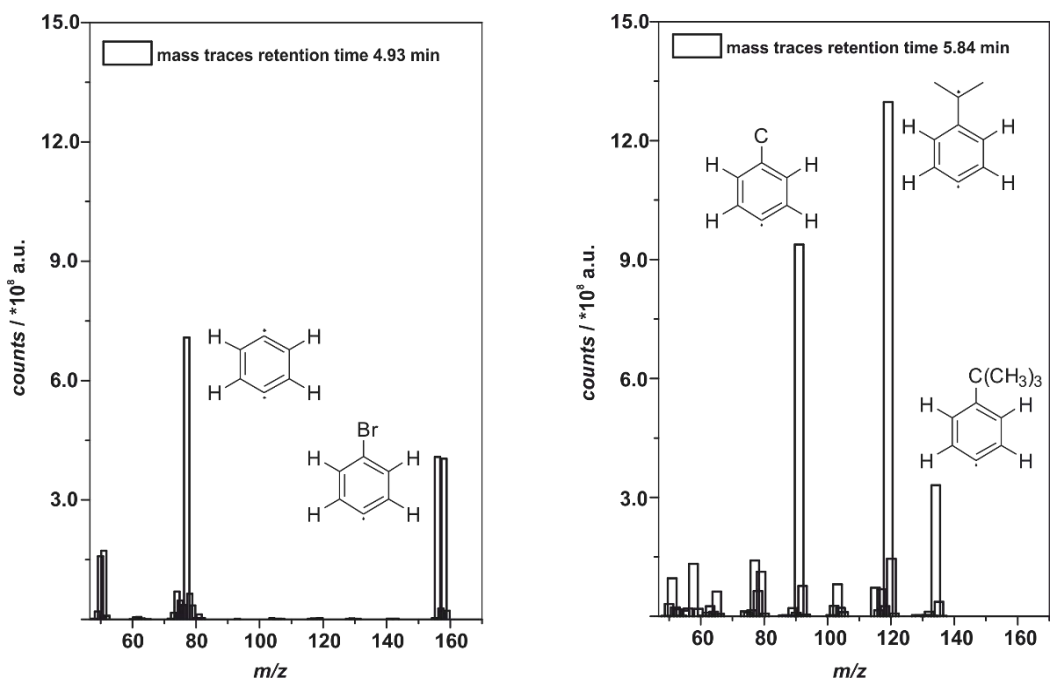

**Figure S15:** MS traces after GC separation obtained at the respective retention times - identification of the detached addends ($T_{TG}$ = 250 °C) of the functionalized carbon allotropes. a) Retention time = 4.93 min: bromobenzene present in **T(1:4)B**, **T(1:4)C**, **G(1:4)B** and **G(1:4)C**. b) Retention time = 5.84 min: *t*-butylbenzene present **T(1:4)A**, **T(1:4)B**, **G(1:4)A** and **G(1:4)B**.

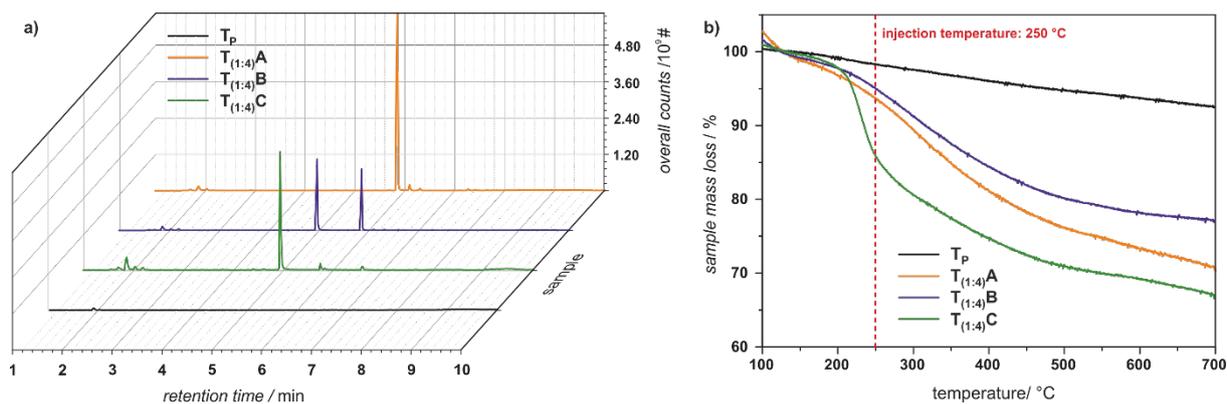

**Figure S16:** a) Elugrams of the detached functional groups of the samples **T$_{(1:4)}$A**, **T$_{(1:4)}$B**, **T$_{(1:4)}$C** and of the pristine SWCNT starting material **T$_P$** at 250 °C – orange: **T$_{(1:4)}$A**, *t*-butylbenzene retention time 5.84; blue: **T$_{(1:4)}$B**, *t*-butylbenzene retention time 5.84 and bromobenzene retention time 4.93; green: **T$_{(1:4)}$C**, bromobenzene retention time 4.93. b) Sample mass loss traces *vs.* temperature in the range of 100-700 °C.

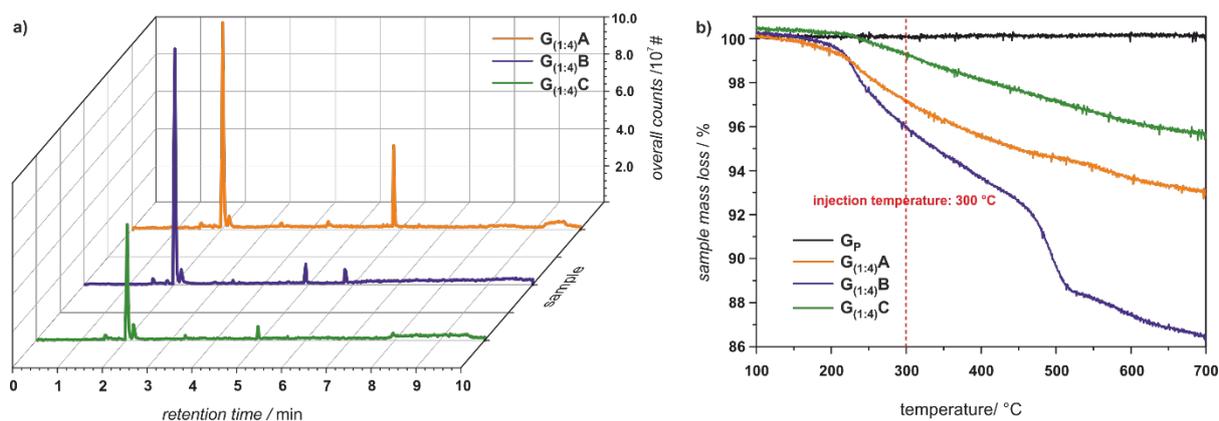

**Figure S17:** a) Elugrams of the detached functional groups of the samples **G$_{(1:4)}$A**, **G$_{(1:4)}$B** and **G$_{(1:4)}$C** at 300 °C – orange: **G$_{(1:4)}$A**, *t*-butylbenzene retention time 5.84; blue: **G$_{(1:4)}$B**, *t*-butylbenzene retention time 5.84 and bromobenzene retention time 4.93; green: **G$_{(1:4)}$C**, bromobenzene retention time 4.93. b) Sample mass loss traces *vs.* temperature in the range of 100-700 °C.

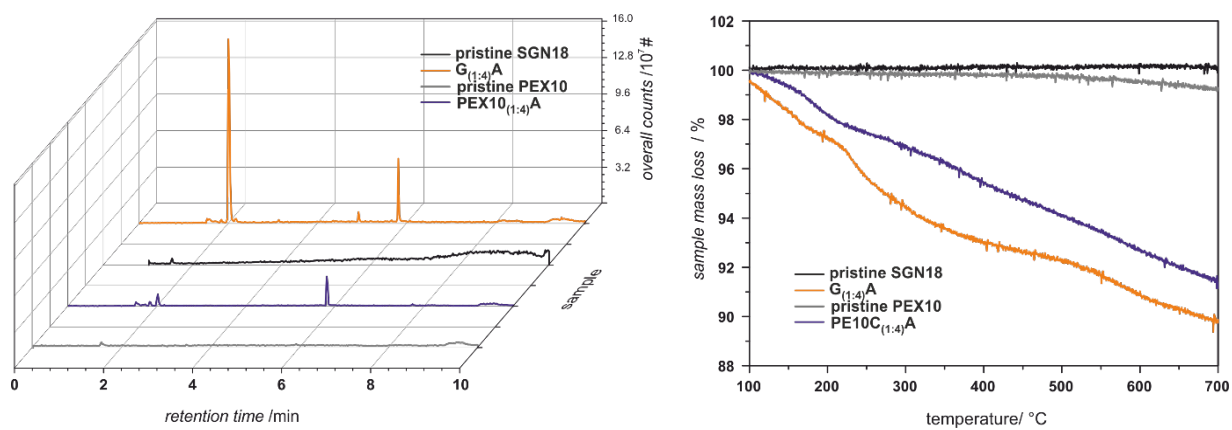

**Figure S18:** a) Reference experiment - variation of flake size. Spherical graphite SGN18 *vs.* powder graphite PEX10. In the case of the starting material with intrinsically smaller flake sizes (PEX10), a negligible amount of THF (RT = 2.04 min) is detected for **PEX$_{(1:4)}$A** by GC.

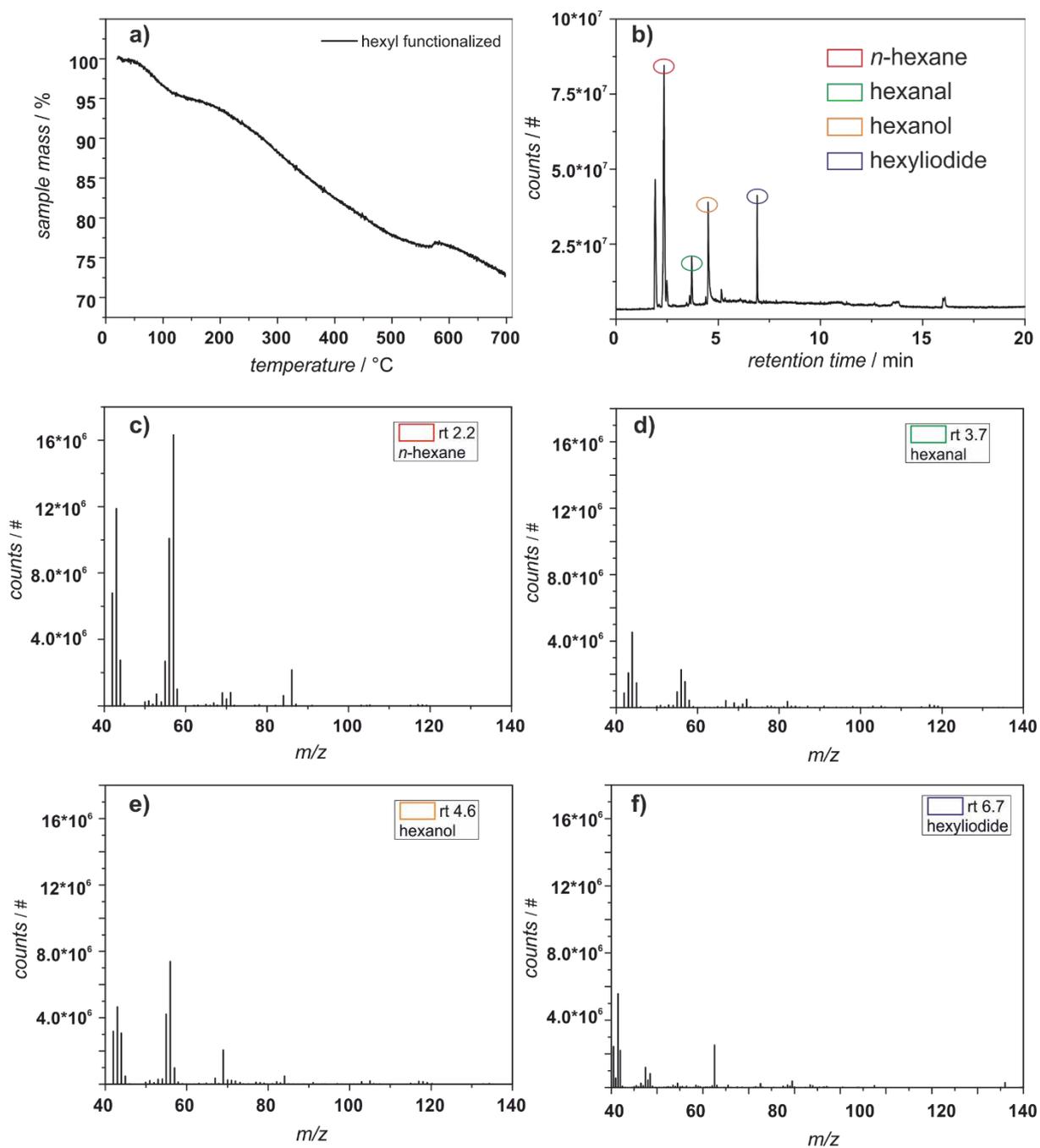

**Figure S19:** TG-GC/MS analysis of a *n*-hexyl-functionalized SWCNT reference sample, obtained *via* reductive activation of the pristine starting material with subsequent trapping of the intermediately formed carbon nanotubides with *n*-hexyliodide.[2] a) Mass loss of the sample *vs.* temperature - 27 % overall mass loss). b) GC-Chromatogram of the temperate – 250 °C – cleaved functional moieties of the *n*-hexyl functionalized sample at 250 °C. c-f) Mass spectroscopic characterization of the GC separated fractions. The respective substances – n-hexane (red), hexanal (green), hexanol (organge), and hexyliodide (blue) were identified on basis of the software integrated NIST MS Search 2.0 modul.

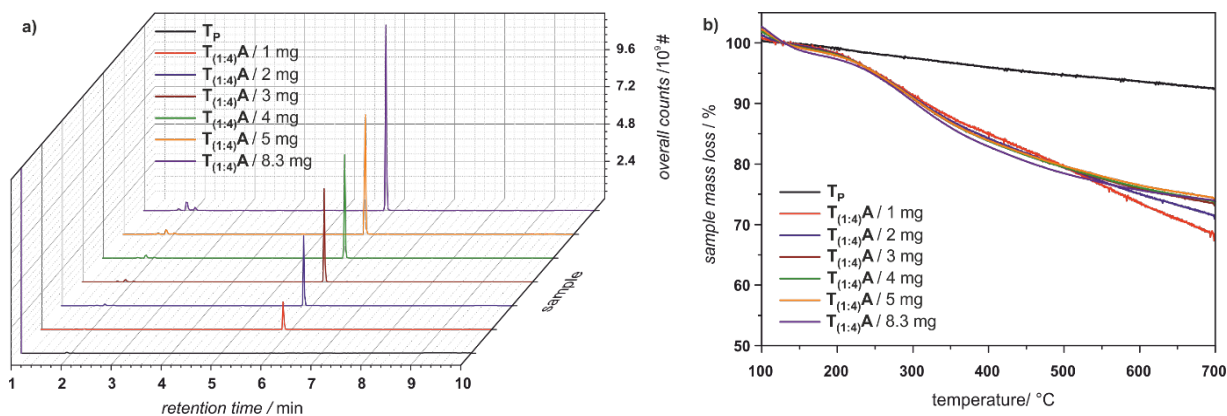

**Figure S20:** a) Elugrams of the detached functional groups ($T_{TG}$ = 250 °C) of the covalently functionalized SWCNT derivative **T$_{(1:4)}$A** obtained for varying weight portions (1-8.3 mg) in the TG. The respective peak area at a retention time of 5.84 min (*t*-butylphenyl) provides the basis for the quantification approach presented in Figure 3 (manuscript). b) Sample mass loss *vs.* temperature in the range of 100-700 °C.

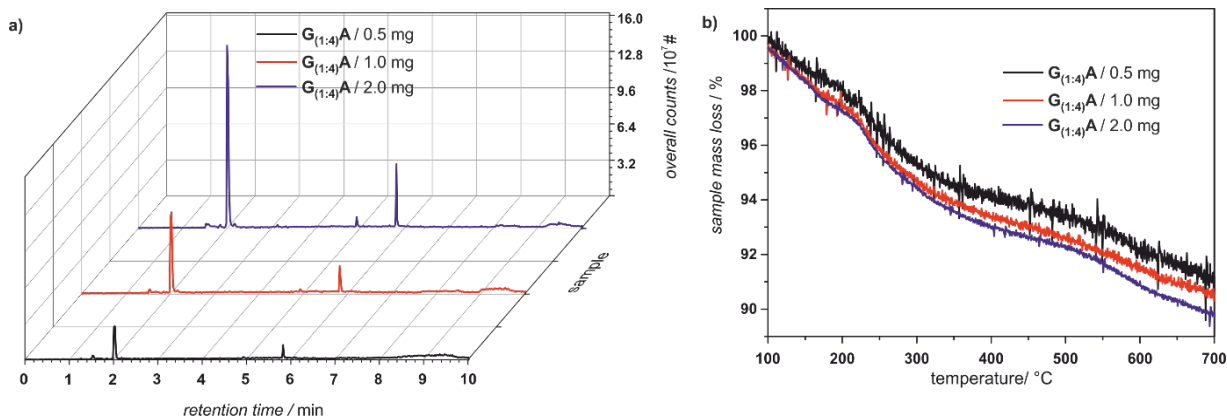

**Figure S21:** a) Elugrams of the detached functional groups ($T_{TG}$ = 300 °C) of the covalently functionalized graphene/graphite derivative **G$_{(1:4)}$A** obtained for varying weight portions (0.5-2 mg) in the TG. The respective peak area at a retention time of 5.84 min (*t*-butylphenyl) provides the basis for the quantification approach presented in Figure 3 (manuscript). b) Sample mass loss *vs.* temperature in the range of 100-700 °C.

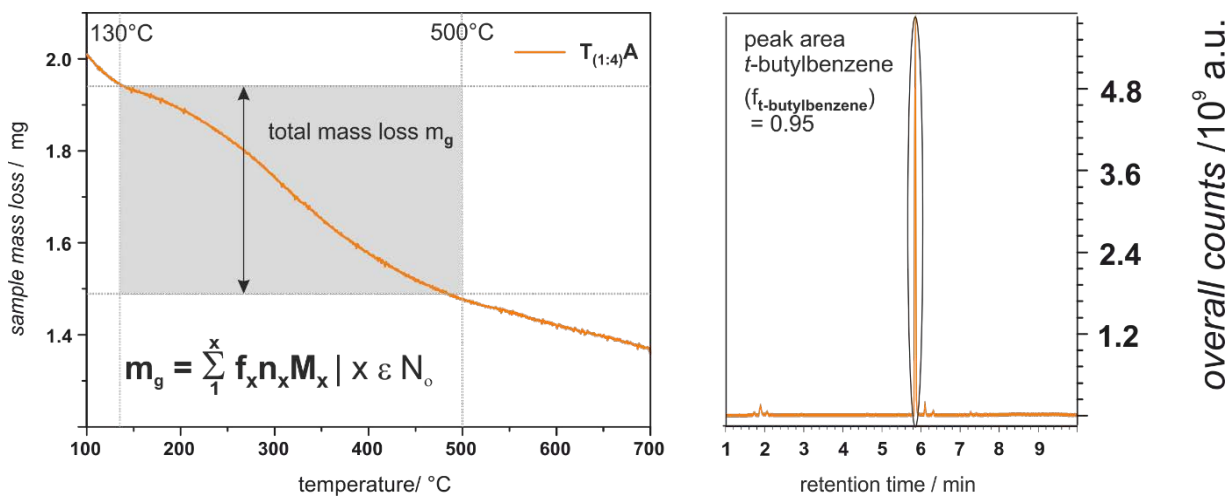

**Figure S22:** The total mass loss ($m_g$), based on the thermal detachment of the *t*-butylphenyl addend, is equal to the sum over the weighting factor $f_x$ (percentage of the respective peak area determined in the GC elugram - right) times the individual mass of the respective component x ($n_x \cdot M_x$ – $n_x$ = molar fraction; $M_x$ = molar mass).

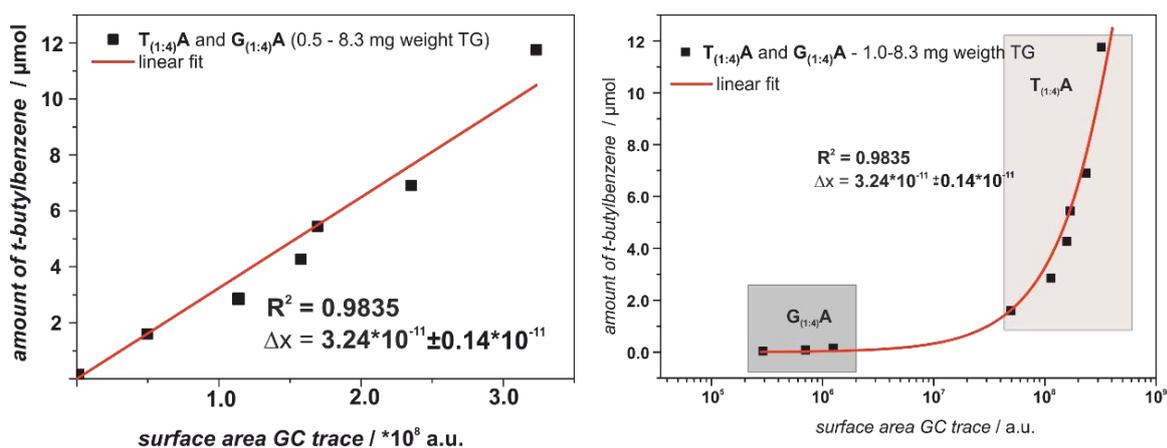

**Figure S23:** Left: TG-GC/MS calibration curve using different sample weights of $T_{(1:4)}A$ and $G_{(1:4)}A$ during the TG measurement. For each TG experiments the corresponding amounts of *t*-butylbenzene are extracted and were correlated to the peak area of the GC-MS elugram. Right: Logarithmic plot of the amount of *t*-butylbenzene attached to covalently functionalized SWCNT – $T_{(1:4)}A$ (1, 2, 3, 4, 5, 8.3 mg) – and graphene/graphite – $G_{(1:4)}A$ (0,5 1, 2 mg) – materials, using different weight portions of samples during the thermogravimetric analysis of the specified samples (see Figure S22 and Figure S23) *vs.* the integrated peak area of the GC-MS trace. The linear fit specifies the factor $3.24 \cdot 10^{-11} \pm 0.14 \cdot 10^{-11}$ and provides the basis for the direct quantification of the functional groups attached to carbon nanotubes as well as to graphene/graphite.

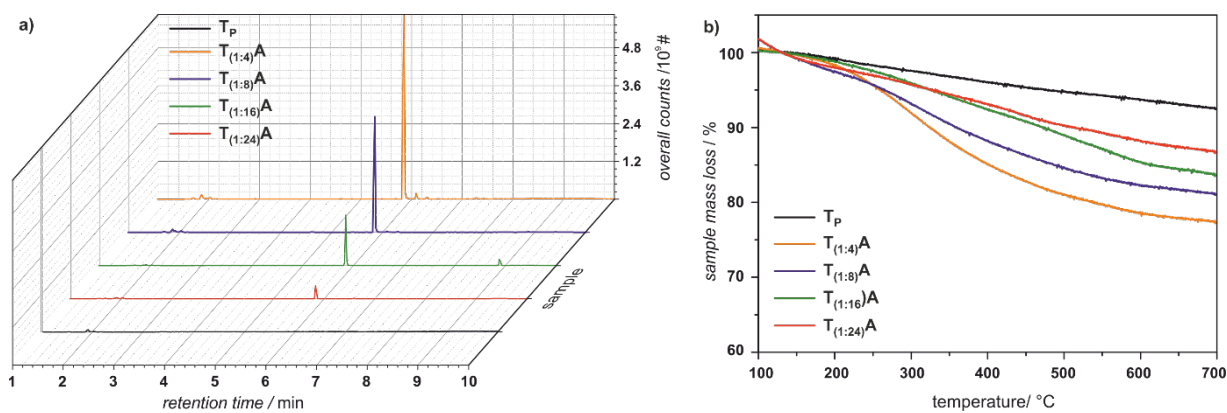

**Figure S24:** a) Elugrams of the detached functional groups ($T_{TG}$ = 250 °C) of $\lambda^3$-iodane **A** functionalized SWCNT derivatives **T$_{(1:4)}$A**, **T$_{(1:8)}$A**, **T$_{(1:16)}$A** and **T$_{(1:24)}$A**. In this case the amount of potassium in the reductive activation step was varied in relation to the respective amount of carbon. The amount of the iodonium salt **A**, used for the trapping of the intermediately charged carbon nanotubides, was kept constant. b) Sample mass loss *vs.* temperature in the range of 100-700 °C.

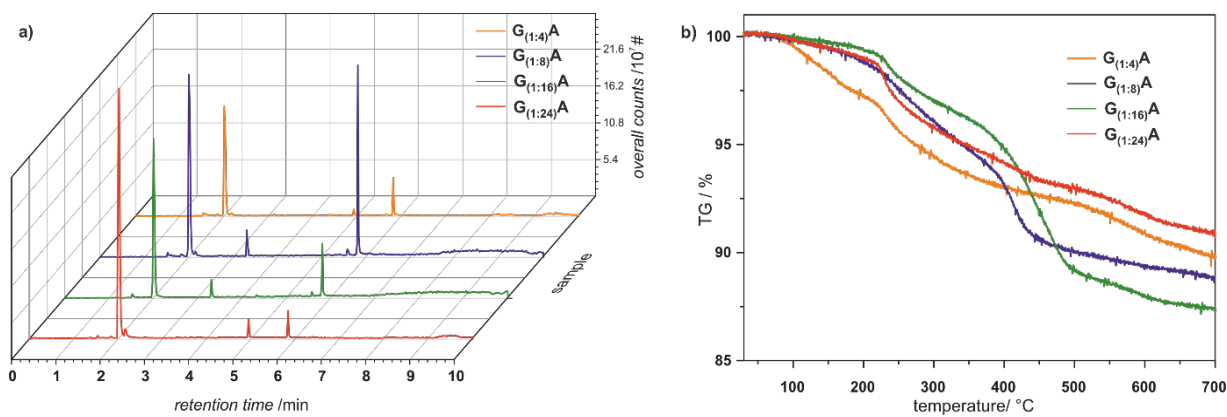

**Figure S25:** a) Elugrams of the detached functional groups ($T_{TG}$ = 300 °C) of $\lambda^3$-iodane **A** functionalized graphene/graphite derivatives **G$_{(1:4)}$A**, **G$_{(1:8)}$A**, **G$_{(1:16)}$A** and **G$_{(1:24)}$A**. In this case the amount of potassium in the reductive activation step was varied in relation to the respective amount of carbon. The amount of the iodonium salt **A**, used for the trapping of the intermediately charged graphenides, was kept constant. b) Sample mass loss *vs.* temperature in the range of 100-700 °C.

**Table ST2:** Quantification of the functional groups, applying the calibration factor determined in Figure 3, varying either the activation parameter for SWCNTs, graphite or the diaryliodonium compound.

| sample | m (Carbon 130 °C-700 °C) / mg | n (addend) / µmol | peak area / counts*$10^6$ | functionalization degree / % |
|---|---|---|---|---|
| T$_{(1:4)}$A | 1.97 | 5.60 | 1726 | 3.42 |
| T$_{(1:4)}$B[a] | 1.91 | 2.03 | 610 | 1.25 |
| T$_{(1:4)}$B[b] | 1.91 | 1.98 | 734 | 1.26 |
| T$_{(1:4)}$C | 1.93 | 3.25 | 1174 | 2.02 |
| T$_{(1:8)}$A | 1.97 | 4.12 | 1269 | 2.45 |
| T$_{(1:16)}$A | 2.01 | 1.30 | 399 | 0.80 |
| T$_{(1:24)}$A | 1.94 | 0.43 | 133 | 0.27 |
| G$_{(1:4)}$A | 2.06 | 0.06 | 18.5 | 0.04 |
| G$_{(1:4)}$B[a] | 2.08 | 0.01 | 3.75 | 0.01 |
| G$_{(1:4)}$B[b] | 2.08 | 0.01 | 3.21 | 0.01 |
| G$_{(1:4)}$C | 2.02 | 0.005 | 1.60 | 0.00 |
| G$_{(1:8)}$A | 2.07 | 0.23 | 71 | 0.13 |
| G$_{(1:16)}$A | 1.98 | 0.07 | 21 | 0.04 |
| G$_{(1:24)}$A | 2.01 | 0.03 | 9.75 | 0.02 |

a) amount of *t*-butylbenzene functionalization, b) bromobenzene functionalization

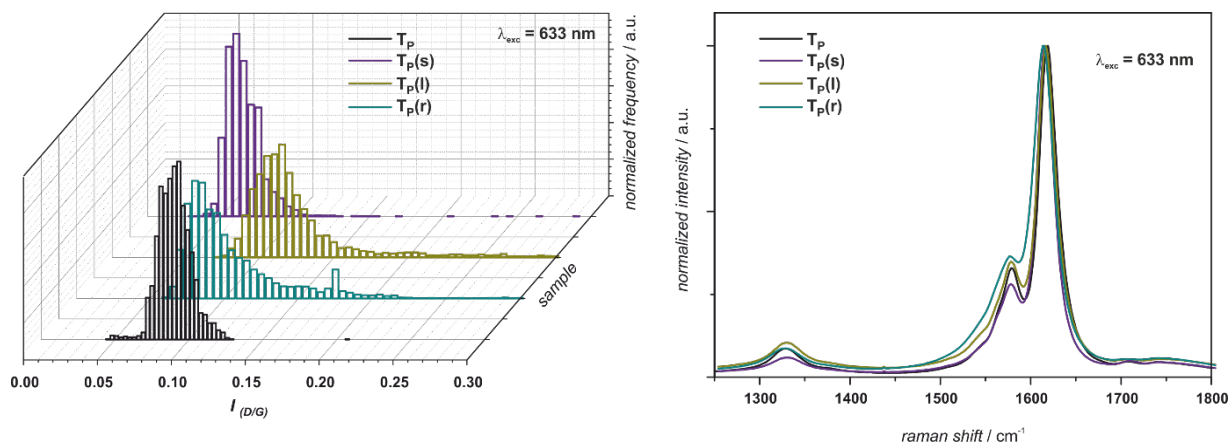

**Figure S26:** Scanning Raman Spectroscopy (SRS) ($\lambda_{exc}$ = 633 nm) of the reference samples **T$_P$(s)** (purple trace), **T$_P$(l)** (yellow trace), and **T$_P$(r)** (teal trace). Here, pristine SWCNT starting material (**T$_P$**) and $\lambda^3$-iodane **A** (*bis*-(4-(t-butyl)phenyl) iodonium triflate) were treated under the following conditions: a) 4 h stirring at RT, yielding **T$_P$(s)**; b) stirring for 4 h under UV light irradiation, yielding **T$_P$(l)**; c) reflux for 4 h, yielding **T$_P$(r)**. Left: Corresponding histograms of the respective I$_{(D/G)}$ values. Right: Mean spectra of the respective SWCNT sample.

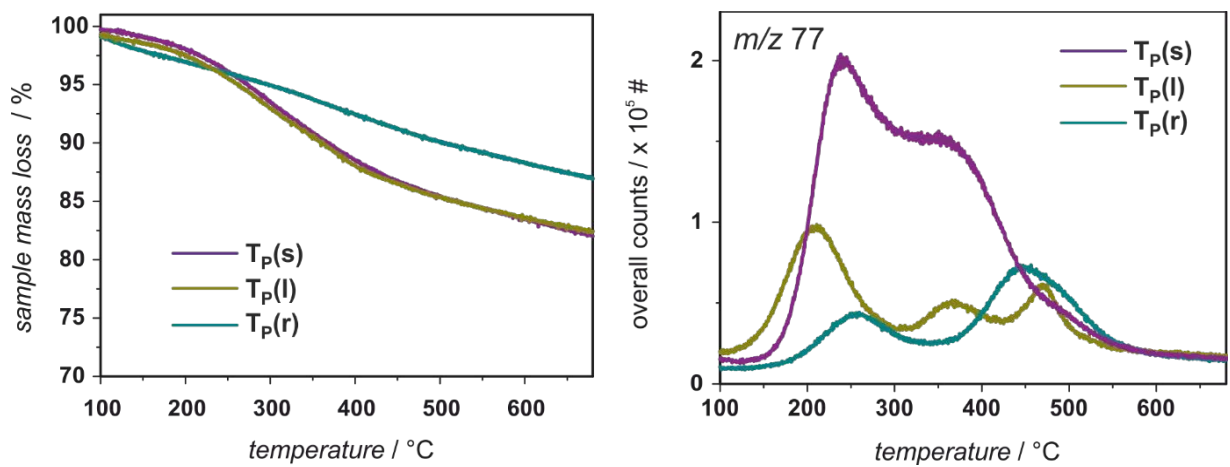

**Figure S27:** TG-MS analysis of the reference samples **T$_P$(s)** (purple trace), **T$_P$(l)** (yellow trace), and **G$_P$(r)** (teal trace). Here, pristine SWCNT starting material (**T$_P$**) and $\lambda^3$-iodane **A** (*bis*-(4-(t-butyl)phenyl) iodonium triflate) were treated under the following conditions: a) 4 h stirring at RT, yielding **T$_P$(s)**; b) stirring for 4 h under UV light irradiation, yielding **T$_P$(l)**; c) reflux for 4 h, yielding **T$_P$(r)**. Left: Corresponding TG-profiles in the temperature range of 100-700 °C. Right: Mass trace of m/z = 77 (C$_6$H$_5$) as a function of TG temperature.

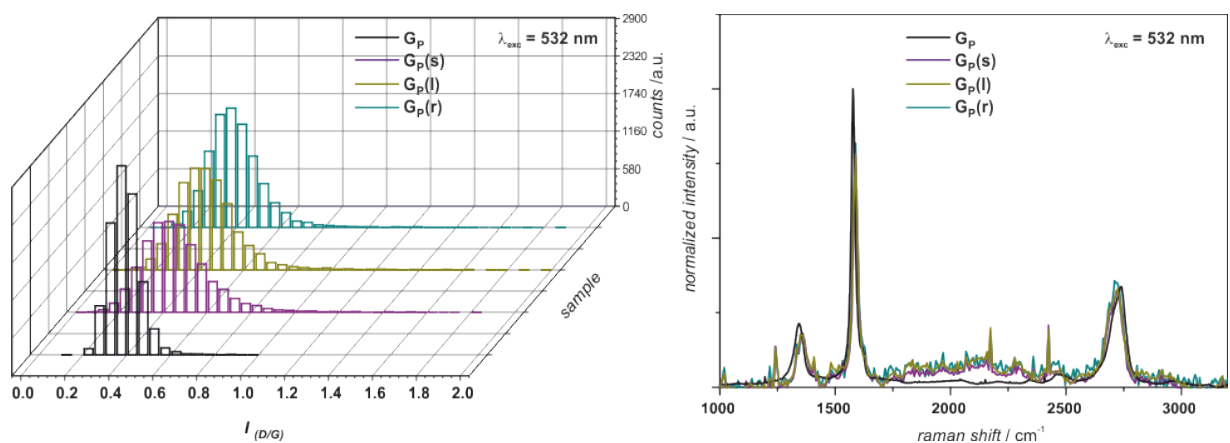

**Figure S28:** Scanning Raman Spectroscopy (SRS) ($\lambda_{exc}$ = 532 nm) of the reference samples **G_P(s)** (purple trace), **G_P(l)** (yellow trace), and **G_P(r)** (teal trace). Here, pristine SGN18 graphite starting material (**G_P**) and $\lambda^3$-iodane **A** (*bis*-(4-(t-butyl)phenyl) iodonium triflate) were treated under the following conditions: a) 4 h stirring, yielding **G_P(s)**; b) stirring for 4 h under UV light irradiation, yielding **G_P(l)**; c) reflux for 4 h, yielding **G_P(r)**. Left: Corresponding histograms of the respective $I_{(D/G)}$ values. Right: Mean spectra of the respective graphene/graphite sample.

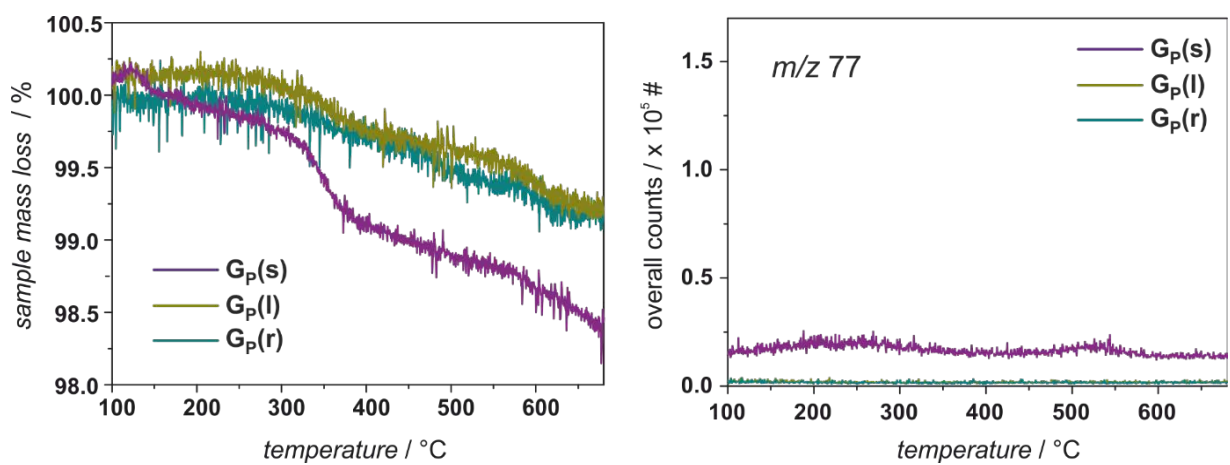

**Figure S29:** TG-MS analysis of the reference samples **G_P(s)** (purple trace), **G_P(l)** (yellow trace), and **G_P(r)** (teal trace). Here, pristine SGN18 graphite starting material (**G_P**) and $\lambda^3$-iodane **A** (*bis*-(4-(t-butyl)phenyl) iodonium triflate) were treated under the following conditions: a) 4 h stirring, yielding **G_P(s)**; b) stirring for 4 h under UV light irradiation, yielding **G_P(l)**; c) reflux for 4 h, yielding **G_P(r)**. Left: Corresponding TG-profiles in the temperature range of 100-700 °C. Right: Mass trace of m/z = 77 ($C_6H_5$) as a function of TG temperature.

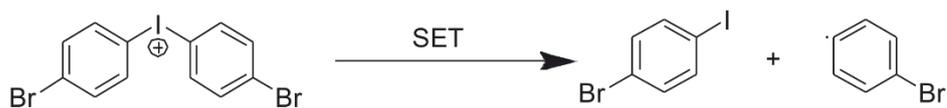

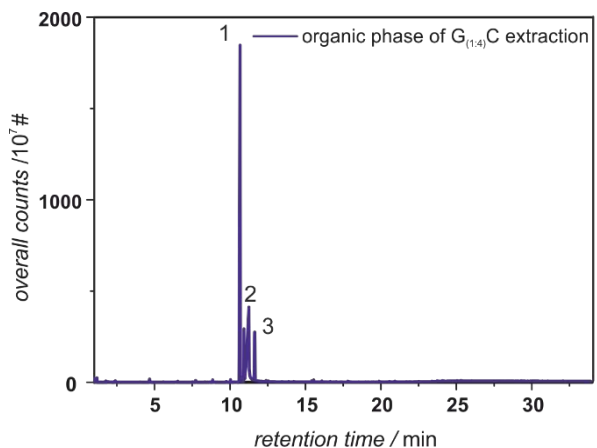
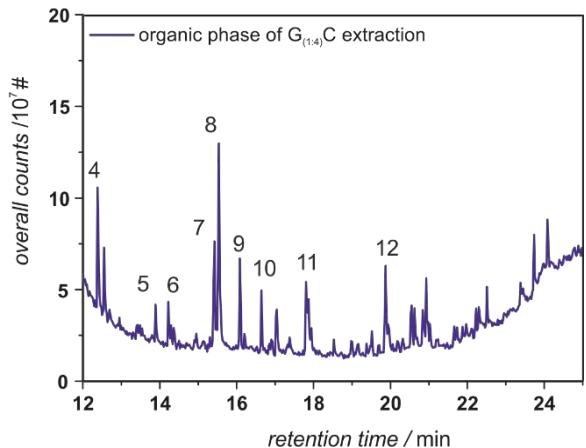

1: benzene, 1-bromo-4-iodo-
2: benzoic acid, 3-chloro- (*)
3: biphenyl
4: benzoic acid, 2-chloro- (*)
5: benzoic acid, 4-chloro- (*)
6: 1,1-biphenyl, 4-bromo-
7: 1,1-biphenyl, 2-iodo-
8: 1,1-biphenyl, 4-bromo-
9: 1,1-biphenyl, 3,4-diethyl
10: acenaphthalene, 1,2-dibromo-
11: o-terphenyl
12: p-terphenyl

**Figure S30:** GC/MS separation and analysis of the different components in the organic extraction phase of **G$_{(1:4)}$C**. 1-Bromo-4-iodobenzene as main component with a retention time of 10.7 min. The characteristic biphenyl derivatives, obtained by radical recombination reactions, are detected at higher retention times. Compounds marked by an asterisk (*) can be attributed to residues from the synthesis of the hypervalent iodane compound **C**.